\documentclass[aps,twocolumn,prx]{revtex4-1}
\usepackage{graphicx,amsmath, amsfonts,bm, color}

\usepackage{xcolor,hyperref}
\hypersetup{
   colorlinks,
   linkcolor={blue!50!black},
   citecolor={blue!50!black},
   urlcolor={blue!80!black}
}

\usepackage{enumitem}

\renewcommand{\=}{\!=\!}
\newcommand{\1}{^{\mbox{\tiny (1)}}}

\DeclareMathOperator{\sgn}{sgn}

\begin{document}

\title{The emergence of crack-like behavior of frictional rupture: The origin of stress drops}
\author{Fabian Barras$^{1}$}
\author{Michael Aldam$^{2}$}
\author{Thibault Roch$^{1}$}
\author{Efim A.~Brener$^{3,4}$}
\author{Eran Bouchbinder$^{2}$}
\thanks{eran.bouchbinder@weizmann.ac.il}
\author{Jean-Fran\c{c}ois Molinari$^{1}$}
\thanks{jean-francois.molinari@epfl.ch}
\affiliation{$^{1}$Civil Engineering Institute, Materials Science and Engineering Institute, Ecole Polytechnique F\'ed\'erale de Lausanne, Station 18, CH-1015 Lausanne, Switzerland\\
$^{2}$Chemical and Biological Physics Department, Weizmann Institute of Science, Rehovot 7610001, Israel\\
$^{3}$Peter Gr\"unberg Institut, Forschungszentrum J\"ulich, D-52425 J\"ulich, Germany\\
$^{4}$Institute for Energy and Climate Research, Forschungszentrum J\"ulich, D-52425 J\"ulich, Germany}

\begin{abstract}
The process of frictional rupture, i.e.~the failure of frictional systems, abounds in the technological and natural world around us, ranging from squealing car brake pads to earthquakes along geological faults. A general framework for understanding and interpreting frictional rupture commonly involves an analogy to ordinary crack propagation, with far-reaching implications for various disciplines, from engineering tribology to geophysics. An important feature of the analogy to cracks is the existence of a reduction in the stress-bearing capacity of the ruptured interface, i.e.~of a drop from the applied stress, realized far ahead of a propagating rupture, to the residual stress left behind it. Yet, how and under what conditions such finite and well-defined stress drops emerge from basic physics are not well understood. Here we show that for rapid rupture a stress drop is directly related to wave radiation from the frictional interface to the bodies surrounding it and to long-range bulk elastodynamics, and not exclusively to the physics of the contact interface. Furthermore, we show that the emergence of a stress drop is a transient effect, affected by the wave travel time in finite systems and by the decay of long-range elastic interactions. Finally, we supplement our results for rapid rupture with predictions for slow rupture. All of the theoretical predictions are supported by available experimental data and by extensive computations. Our findings elucidate the origin of stress drops in frictional rupture, i.e.~they offer a comprehensive and fundamental understanding of why, how and to what extent frictional rupture might be viewed as an ordinary fracture process.
\end{abstract}

\maketitle

\section{Background and motivation}
\label{sec:intro}

Rapid slip along interfaces separating bodies in frictional contact is mediated by the spatiotemporal dynamics of frictional rupture~\cite{Svetlizky2019,Scholz2002}. Frictional rupture is a fundamental process of prime importance for a broad range of physical systems, e.g.~it is responsible for squealing in car brake pads~\cite{Rhee1991}, for bowing on a violin string~\cite{Casado2017}, and for earthquakes along geological faults~\cite{Marone1998a,Ben-Zion2008,Ohnaka2013}, to name just a few well-known examples. The essence of frictional rupture propagation is that a state of relatively high slip rate (the rate of interfacial shear displacement discontinuity) behind the rupture edge propagates into a low/vanishing slip rate state ahead of it, cf.~Fig.~\ref{fig:Fig1}. As such, frictional rupture appears to be essentially similar to ordinary tensile (opening) cracks, where a finite tensile displacement discontinuity (broken material) state behind the crack edge propagates into a zero tensile displacement discontinuity (intact material) state ahead of it~\cite{Freund1998}.

There is, however, an important and fundamental difference between frictional rupture and ordinary tensile cracks that manifests itself in the stress states associated with these two processes. A tensile crack, i.e.~a crack subjected to opening forces, is composed of surfaces that cannot support stress, so the stress behind its edge vanishes. Consequently, tensile (opening) crack propagation is a process in which the applied stress ahead of the crack edge drops to zero behind it. This stress drop, which accompanies tensile crack propagation, has dramatic implications. Most notably, the loss of stress-bearing capacity along the crack surfaces is compensated by large concentration of deformation and stress near the crack edge, oftentimes in a way that mimics a mathematical singularity, whose intensity increases with increasing stress drops~\cite{Freund1998}. Frictional rupture is different from tensile cracks because the finite frictional interaction between the two bodies in contact behind the rupture edge generically implies that the stress there cannot drop to zero, but rather remains finite.

The close relations between frictional rupture and tensile cracks can be maintained if, as is widely assumed, the stress behind the frictional rupture edge --- the residual stress $\tau_{\rm res}$ --- is well-defined and is generically smaller than the far-field stress $\tau_{\rm d}$ that is required to drive rupture. Moreover, the residual stress $\tau_{\rm res}$ is generally assumed to be an intrinsic interfacial property of the slipping contact interface, typically related to the kinetic friction coefficient. Under these assumptions, a finite stress drop $\Delta\tau\!\equiv\!\tau_{\rm d}-\tau_{\rm res}\!>\!0$ exists and effective crack-like properties of frictional rupture, e.g.~edge singularity, are expected to emerge. These assumptions have been adopted in an extremely broad range of theoretical and numerical studies~\cite{Ida1972,Palmer1973,Andrews1976,Madariaga1977,Cochard2000,Bizzarri2001,Uenishi2003,Rice2005,Bhat2007,Liu2008,Dunham2008,Bizzarri2010,Bizzarri2010a,Viesca2012,Kammer2012,Liu2014,Kammer2015, Bizzarri2016,Barras2017}, and their implications have been consistent with geophysical observations~\cite{Abercrombie2005,Bizzarri2016} and have been confirmed in some recent laboratory experiments~\cite{Lu2010,Lu2010a,Noda2013a,Svetlizky2014,Bayart2015,Svetlizky2016,Rubino2017,Svetlizky2017a}. In fact, stress drops are among the few remotely observable parameters in earthquake science, providing a key link to the frictional properties of faults, which can be only indirectly inferred.

Yet, to the best of our knowledge, currently there is no basic understanding of how and under what conditions the effective crack-like behavior of frictional rupture emerges from fundamental physics. More specifically, there is a need to understand what the physical origin of a finite stress drop $\Delta\tau\!>\!0$ is and under what conditions it emerges. Here we address these basic questions; first, we show that for rapid rupture a finite and well-defined stress drop is not an interfacial property, as is widely assumed, but rather it is directly related to wave radiation from the frictional interface to the bodies surrounding it (the so-called radiation damping effect~\cite{Ben-Zion1995,Perrin1995,Zheng1998,Crupi2013}) and to long-range bulk elastodynamic interaction effects. Second, we show that the emergence of a stress drop is a finite time effect, limited by the wave travel time in finite systems. Third, we show that for slow rupture, i.e.~rupture that is significantly slower than the elastic wave-speeds~\cite{Peng2010,Obara2016,Takagi2016,Gomberg2016}, stress drops are transiently controlled by the long-range quasi-static elasticity of the bodies surrounding the frictional interface. Reanalysis of very recent experimental results, reported by two different experimental groups, provides strong support to our theoretical predictions, for both rapid and slow rupture. All in all, our findings elucidate the origin of stress drops in frictional rupture, i.e.~they offer a comprehensive and fundamental understanding of why, how and to what extent frictional rupture might be viewed as an ordinary fracture process.

\section{The physical origin and magnitude of the stress drop associated with frictional rupture}
\label{sec:SD}

The starting point for our discussion is a physically-motivated interfacial constitutive law, i.e.~a relation between the dynamical and structural variables that characterize a frictional interface and the frictional resistance stress $\tau$~\cite{Baumberger2006}. A frictional interface is formed when two bodies come into contact. Each of them satisfies its own continuum momentum balance equation $\rho\ddot{\bm u}(\bm r, t)\=\nabla\!\cdot{\bm \sigma}(\bm r, t)$, where $\rho$ is the mass density, $\bm u$ and $\bm r\=(x,y)$ (in two-dimensions) are the displacement and position vector fields respectively, and $\bm \sigma$ is the stress tensor field (a superposed dot represents a time derivative). $\bm \sigma$ in each body is related to $\bm u$ through a bulk constitutive law (i.e.~a constitutive law that characterizes the bodies forming the interface), oftentimes Hooke's law of linear elasticity, to be adopted below as well. Note that body forces are neglected in the momentum balance equation.

The interfacial constitutive law involves three bulk quantities evaluated at the interface located at $y\=0$: (i) the slip rate/velocity $v(x,t)\!\equiv\!\dot{u}_x(x,y\=0^+,t)\!-\!\dot{u}_x(x,y\=0^-,t)$, where $+/-$ correspond to the upper/lower bodies, respectively (ii) the shear stress $\sigma_{xy}(x,y\=0,t)$ that is balanced by the frictional stress, $\tau(x,t)\=\sigma_{xy}(x,y\=0,t)$ and (iii) the normal stress $\sigma(x,t)\!\equiv\!-\sigma_{yy}(x,y\=0,t)$. A large body of evidence accumulated in the last few decades indicates that the interfacial constitutive law must also involve a set of non-equilibrium order parameters $\{\phi_i\}$, sometimes termed internal-state fields, that represent the structural state of the interface and encode its history~\cite{Ruina1983,Rice1983,Marone1998,Nakatani2001,Baumberger2006,Dietrich2007,Nagata2012,Bhattacharya2014}. In a minimal formulation, adopted in numerous studies~\cite{Dieterich1992,Perrin1995,Roy1996,Ben-Zion1997,Zheng1998,Baumberger1999,Lapusta2000,Aldam2017a}, a single internal-state field $\phi(x,t)$ is used. This assumption is adopted here, without loss of generality.

The interfacial constitutive law, at any position $x$ along the interface and at any time $t$, is described by the following local relation
\begin{equation}
\label{eq:friction_law}
\tau=\sigma\,\sgn(v)\,f(|v|,\phi) \ ,
\end{equation}
which must be supplemented with a dynamical equation for the evolution of $\phi$. Extensive evidence indicates that $\phi$ physically represents the age/maturity of the contact~\cite{Rice1983,Marone1998,Nakatani2001,Baumberger2006,Dietrich2007,Nagata2012,Bhattacharya2014} and that its evolution takes the form
\begin{equation}
\label{eq:dot_phi}
\dot\phi = g\left(\frac{|v|\phi}{D}\right) \ ,
\end{equation}
with $g(1)\=0$ and where $\phi$ is of time dimension. The characteristic slip displacement $D$ controls the transition from a stick state $v\!\approx\!0$, with a characteristic structural state $\phi\=\phi_0$, to a steadily slipping/sliding state $v\!>\!0$, with $\phi_{\rm ss}\=D/v$. The precise functional form of $g(\cdot)$ (with $g(1)\=0$) plays no role in what follows.

The function $f(|v|,\phi_{\rm ss}\=D/v)\=\tau_{\rm ss}(v)/\sigma$, under steady-state sliding conditions and a controlled normal stress $\sigma$, has been measured over a broad range of slip rates $v$ for many materials~\cite{Baumberger2006}. Together with general theoretical considerations~\cite{Bar-Sinai2014}, it is now established that the steady-state frictional stress $\tau_{\rm ss}(v)$ is generically $N$-shaped, as shown in Fig.~\ref{fig:Fig2}a. Consider then a frictional system driven by a shear stress $\tau_{\rm d}$, which is larger than the minimum of the $\tau_{\rm ss}(v)$ curve, cf.~Fig.~\ref{fig:Fig2}a. What are the generic properties of frictional rupture that might emerge under these conditions?
\begin{figure}[ht!]
  \centering
  \includegraphics[width=\columnwidth]{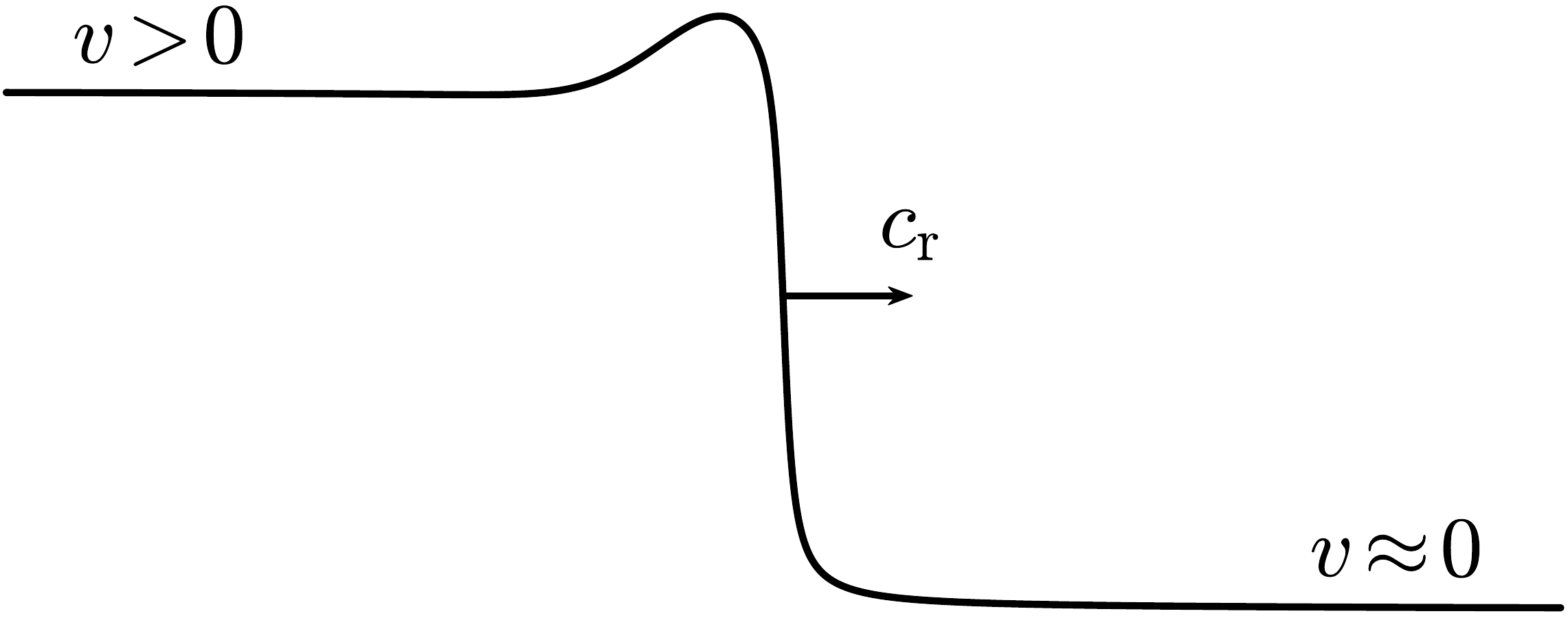}
  \caption{A schematic representation of the spatial slip rate $v$ profile of a frictional rupture propagating at a velocity $c_{\rm r}$ from left to right. A slipping/sliding state with a relatively high slip velocity, $v\!>\!0$, characterizes the interface behind the propagating rupture edge and a low/vanishing slip rate state, $v\!\approx\!0$, characterizes the interface ahead of it.}\label{fig:Fig1}
\end{figure}

As explained in Sect.~\ref{sec:intro}, frictional rupture is a propagating spatiotemporal object that features a relatively high slip state $v\!>\!0$ behind its edge and a stick (no/slow slip) $v\!\approx\!0$ state ahead of it, as shown in Fig.~\ref{fig:Fig1}. This spatiotemporal dynamical process can be directly related to Fig.~\ref{fig:Fig2}a, where the driving stress $\tau_{\rm d}$ intersects the $\tau_{\rm ss}(v)$ curve at three points. The leftmost intersection point features an extremely small slip velocity, $v\!\approx\!0$, which corresponds to the state ahead of the rupture edge. We stress that in general the state of the interface ahead of the rupture edge may be far from steady-state and the results to follow are largely independent of its detailed properties. The rightmost intersection point features a relatively large slip velocity, $v\!>\!0$, which corresponds to the state behind the rupture edge. The transition between these two states takes place in the edge region, and is controlled by $\dot\phi$ in Eq.~\eqref{eq:dot_phi} and by spatiotemporal bulk dynamics. In this transition region, the slip velocity $v$ also goes through the intermediate intersection point, which is not a stable fixed point as the other two. The crucial observation is that the stress behind steadily propagating frictional rupture is $\tau_{\rm d}$, i.e.~the residual stress equals to the driving stress, $\tau_{\rm res}\=\tau_{\rm d}$. This implies that we expect {\em no stress drop to emerge at all}, $\Delta\tau\=\tau_{\rm d}-\tau_{\rm res}\=0$, and consequently no crack-like behavior.

In many studies available in the literature, a steady-state friction curve $\tau_{\rm ss}(v)$ that does not feature a minimum is adopted~\cite{Perrin1995,Zheng1998,Rubin2005,Ampuero2008,Nielsen2017}. We consequently discuss here such a no-minimum friction curve and plot an example of it in Fig.~\ref{fig:Fig2}a. In this case, the driving stress $\tau_{\rm d}$ behind the rupture edge cannot be balanced by the friction stress and the slip velocity in this region is expected to continuously accelerate. As such, we cannot expect a well-defined steady-state stress drop to emerge, though the stress will definitely drop below $\tau_{\rm d}$ behind the rupture edge.

The discussion above, both for the $N$-shaped and the no-minimum steady-state friction curves, seems to lead to the quite remarkable conclusion that based on basic physics considerations we expect no finite and well-defined steady-state stress drops to emerge at all in the context of friction rupture, and hence no crack-like behavior as well. This appears to be in sharp contrast to ample evidence indicating the existence of finite and well-defined steady-state stress drops in various frictional systems~\cite{Lu2010,Lu2010a,Svetlizky2014,Bayart2015,Bayart2016,Svetlizky2017,Rubino2017,Svetlizky2019}. How can one reconcile the two apparently conflicting conclusions?

To address this question, let us write down in more detail the general expression for interfacial shear stress $\sigma_{xy}(x,y\=0,t)$, valid also out of steady-state, when a constant driving stress $\tau_{\rm d}$ is applied at the far boundaries of the systems, say at $y\=\pm H$ ($H$ is the height of each of the two bodies in contact). For bulk linear elastodynamics, we have $\sigma_{xy}(x,y\=0,t)\=\tau_{\rm d}+\tilde{s}(x,t)$, where $\tilde{s}(x,t)$ is a spatiotemporal integral that quantifies the long-range (in both space and time) elastodynamic interaction between different parts of the interface~\cite{Geubelle1995,Morrissey1997,Breitenfeld1998}. Under strict homogeneous steady-state conditions, we have $\tilde{s}(x,t)\!\to\!0$ and consequently $\sigma_{xy}(x,y\=0,t)\to\tau_{\rm d}$, which corresponds to the rightmost intersection point Fig.~\ref{fig:Fig2}a attained far behind the rupture edge (see discussion above). At finite times, before strict steady-state conditions are attained, the spatiotemporal integral term $\tilde{s}(x,t)$ makes a finite contribution to $\sigma_{xy}(x,y\=0,t)$, which quantifies the deviation from steady-state.

Under these conditions, and in particular for times in which information regarding the evolution of the slip velocity $v(x,t)$ relative to some initial/reference slip velocity $v_0$ does not have enough time to propagate to the boundaries at $y\=\pm H$ and back to the interface, the spatiotemporal integral term $\tilde{s}(x,t)$ can be decomposed into two contributions, one is a local contribution of the form $\frac{\mu}{2c_s}\!\left(v(x,t)-v_0\right)$, where $\mu$ is the linear elastic shear modulus and $c_s$ is the shear wave-speed, and the other is a non-local (in space and time) contribution $s(x,t)$~\cite{Geubelle1995,Morrissey1997,Breitenfeld1998}. This decomposition is valid for times shorter than ${\cal O}(H/c_s)$, before wave interaction with the boundaries is possible, and for these times the interfacial shear stress takes the form~\cite{Geubelle1995,Rice1996,Ben-Zion1997,Morrissey1997,Breitenfeld1998,Lapusta2000}
\begin{equation}
\label{eq:RD}
\sigma_{xy}(x,y\=0,t)=\tau_{\rm d}-\frac{\mu}{2c_s}\Big(v(x,t)-v_0\Big) + s(x,t) \ .
\end{equation}
In many studies available in the literature the idealized infinite system limit $H\!\to\!\infty$ is considered, for which Eq.~\eqref{eq:RD} is valid at all times. The term $\frac{\mu}{2c_s}\!\left(v(x,t)-v_0\right)$ physically represents wave radiation from the interface to the bodies that form it and is therefore known as the radiation damping term~\cite{Ben-Zion1995,Perrin1995,Zheng1998,Crupi2013}. It is associated with ``damping'' because from the perspective of the interface it acts as a viscous stress with $\mu/2c_s$ being the effective viscosity. This term makes an important contribution to stress drops in frictional rupture, as is shown next.

Consider a point along a frictional interface that is initially located ahead of a propagating frictional rupture and whose slip velocity is $v\!\approx\!0$, which represents $v_0$ in Eq.~\eqref{eq:RD}. When the frictional rupture goes through this point, the stress and slip velocities vary significantly. Suppose then that the system height $H$ is sufficiently large such that the spatiotemporal integral decays once the rupture went sufficiently far ahead, $s(x,t)\!\to\!0$, but the radiation damping contribution is still valid (i.e.~shear waves did not have enough time to propagate to the far boundaries and back). Under these conditions, the slip velocity $v_{\rm res}$ at the spatial point under consideration, now far behind the frictional rupture, is determined by the shear stress balance $\sigma_{xy}\=\tau$ and satisfies
\begin{equation}
\label{eq:v_r}
\tau_{\rm ss}(v^0_{\rm res})+\frac{\mu}{2c_s}v^0_{\rm res}\simeq \tau_{\rm d} \ ,
\end{equation}
where $v^0_{\rm res}\!\gg\!v_0$ and $s(x,t)\!\to\!0$ have been used. Note that the superscript '$0$' in $v^0_{\rm res}$ represents the fact that this is the theoretically predicted residual slip velocity under the assumption that $s(x,t)\=0$ far behind the propagating rupture front. The residual stress at this point takes the form $\tau_{\rm res}\=\tau_{\rm d}-\frac{\mu}{2c_s}v^0_{\rm res}$, and consequently a finite stress drop of magnitude
\begin{equation}
\label{eq:SD}
\Delta\tau \simeq \frac{\mu}{2c_s}v^0_{\rm res} \ ,
\end{equation}
is expected to emerge on times shorter than ${\cal O}(H/c_s)$.

A geometric representation of Eqs.~\eqref{eq:v_r}-\eqref{eq:SD} is shown in Fig.~\ref{fig:Fig2}b for the $N$-shaped $\tau_{\rm ss}(v)$ and in Fig.~\ref{fig:Fig2}c for the no-minimum $\tau_{\rm ss}(v)$, both shown in Fig.~\ref{fig:Fig2}a. In Figs.~\ref{fig:Fig2}b-c, the left-hand-side of Eq.~\eqref{eq:v_r} $\tau_{\rm ss}(v)+\frac{\mu}{2c_s}v$ is regarded as an {\em effective} steady-state curve and is plotted by a dashed line. The radiation damping contribution $\frac{\mu}{2c_s}v$ in Fig.~\ref{fig:Fig2}b shifts the location of the effective (finite time) steady-state slip rate to lower rates (compared to the strict steady-state represented by the rightmost intersection point in Fig.~\ref{fig:Fig2}a). In Fig.~\ref{fig:Fig2}c it gives rise to an effective (finite time) steady-state slip rate, which simply does not exist for the no-minimum curve in Fig.~\ref{fig:Fig2}a. This shows that Eq.~\eqref{eq:SD} is valid independently of the properties of $\tau_{\rm ss}(v)$. Note that relevant solutions $v^0_{\rm res}$ to Eq.~\eqref{eq:v_r} exist only if $\tau_{\rm d}$ is larger than the effective (finite time) minimum of the steady-state friction curve, as is highlighted in Figs.~\ref{fig:Fig2}b-c. The very same condition played a central role in the analysis of~\cite{Zheng1998}, where the conditions for rupture mode selection (self-healing pulses vs.~crack-like) have been extensively discussed. Somewhat related issues have also been discussed in~\cite{Cochard1994}.
\begin{figure}[ht!]
  \centering
  \includegraphics[width=\columnwidth]{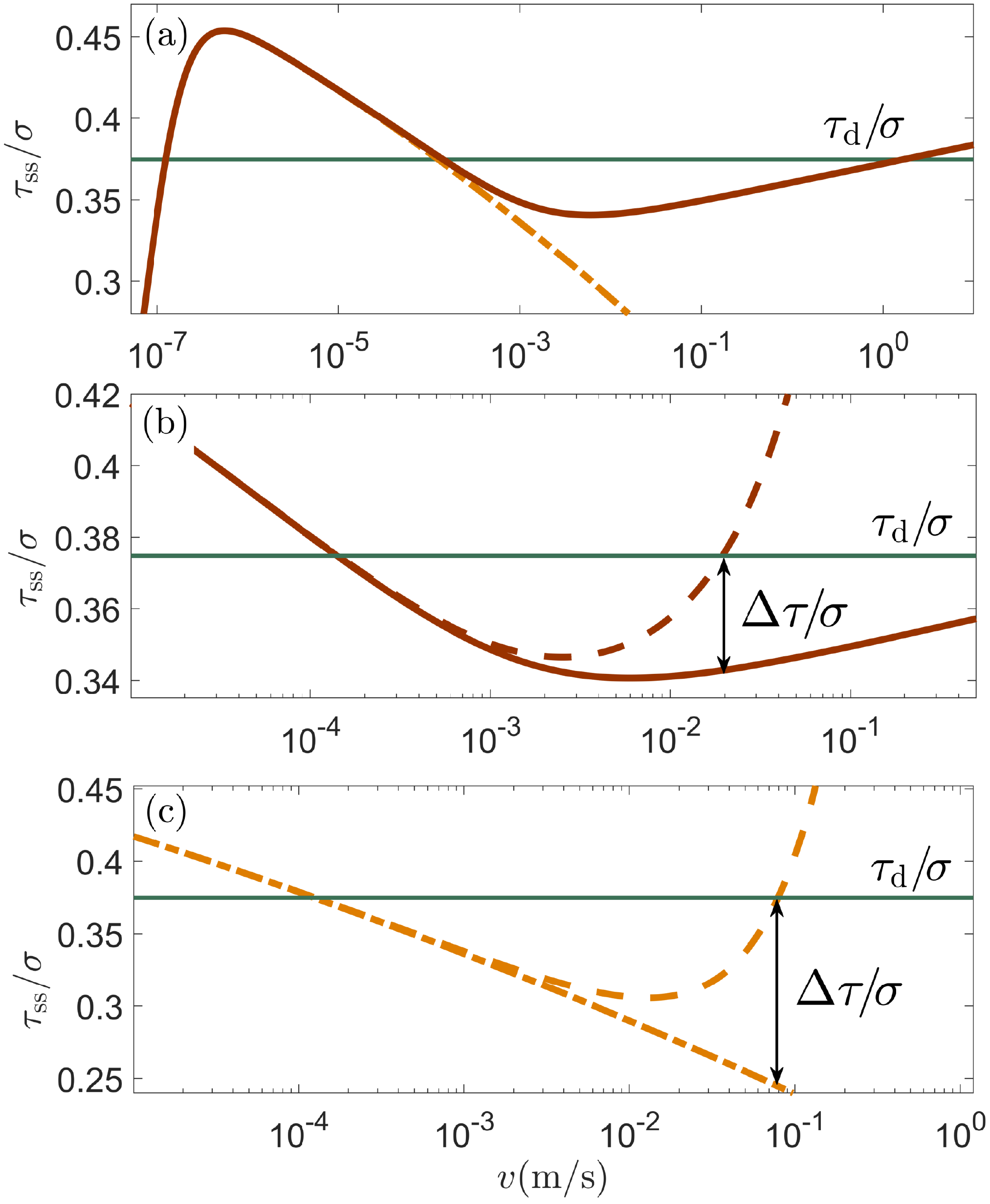}
  \caption{(a) A schematic representation of the steady-state friction stress $\tau_{\rm ss}(v)$, normalized by a constant normal stress $\sigma$, vs.~the slip rate $v$ (solid brown line). The curve has a generic $N$-shape~\cite{Bar-Sinai2014}, with a maximum at an extremely low $v$ and a minimum at an intermediate $v$. The horizontal line represents the driving stress $\tau_{\rm d}$, which intersects the $N$-shaped steady-state friction curve at three points; the leftmost and rightmost ones are stable fixed points, while the intermediate one is an unstable one. Also shown is a steady-state friction curve without the minimum (dash-dotted orange line)~\cite{Ruina1983,Marone1990,Scholz1998}. This no-minimum steady-state friction curve lacks the rightmost intersection point of the solid brown curve. (b) The effective steady-state friction curve (dashed brown line), obtained by adding $\frac{\mu}{2c_s}v$ (with $\mu\!=\!9$GPa and $c_s\!=\!2739$m/s) to the solid brown line of panel (a), together with a copy of the solid brown line of panel (a) itself. The intersection of the dashed brown line with the horizontal $\tau_{\rm d}$ line (the same as in panel (a)) is described by Eq.~\eqref{eq:v_r} and the stress drop $\Delta\tau$ of Eq.~\eqref{eq:SD} is marked by the black double-arrow. (c) The same as panel (b), but for the no-minimum curve (dash-dotted orange line) of panel (a).}\label{fig:Fig2}
\end{figure}

We would like to note that once $\Delta\tau$ is {\em assumed} to exist, the relation in Eq.~\eqref{eq:SD} between $\Delta\tau$ and $v_{\rm res}$ as {\em given quantities}, is a known elastodynamic relation~\cite{Brune1970}, which has previously received some experimental support (e.g.~see Fig.~8 in~\cite{Okubo1984}) and is strongly supported by recent experimental data we have extracted from recent experimental work~\cite{Rubino2017, Svetlizky2017a}, see Sect.~\ref{sec:exp}. To the best of our knowledge, however, none of these works (except for~\cite{Zheng1998}, as mentioned above) predicted the residual slip velocity $v_{\rm res}$ to emerge from the combined effect of the steady-state friction curve $\tau_{\rm ss}(v)$ and of the radiation damping term $\frac{\mu}{2c_s}v$ as predicted in Eq.~\eqref{eq:v_r}, and none of them interpreted Eq.~\eqref{eq:SD} as a major contribution to stress drops $\Delta\tau$ in frictional rupture. In the context of a quasi-static fault model of~\cite{Rice1993}, the radiation damping term $\frac{\mu}{2c_s}v$ has been added in an ad hoc manner in order to avoid unbounded slip velocities to emerge during frictional instabilities in the quasi-static formulation. It has been noted in this context~\cite{Rice1993} that the effective viscosity $\mu/2c_s$ affects the magnitude of the stress drop $\Delta\tau$ when $\tau_{\rm ss}(v)$ has no-minimum (cf.~Fig.~\ref{fig:Fig2}c); nevertheless, the general physical picture in which the radiation damping term significantly contributes to stress drops in frictional rupture, independently of the properties of $\tau_{\rm ss}(v)$, has not been discussed.

When the radiation damping term $\tfrac{\mu}{2c_s}v$ in Eq.~\eqref{eq:v_r} does not faithfully represent the physics of a given system, no well-defined finite stress drop is expected to accompany rapid frictional rupture. This can happen in two generic cases; first, in the limit of thin bodies $H\!\to\!0$, where essentially there is no bulk to radiate energy into, the radiation damping term $\tfrac{\mu}{2c_s}v$ simply does not exist to begin with. In this case, frictional rupture exists, but it is not accompanied by any stress drop, as shown previously in~\cite{Bar-Sinai2012} and here in~Fig.~A1. Second, the radiation damping term $\tfrac{\mu}{2c_s}v$, which exists at relatively short times (shorter than ${\cal O}(H/c_s)$), is expected to vanish in the long time limit $t\!\gg\!H/c_s$. This limit can be probed by performing experiments or simulations with long enough systems for long enough times. Indeed, simulations of effectively long systems yield rupture fronts with no stress drop, see Fig.~6a in~\cite{feature} and details therein. Finally, note that the radiation damping term $\tfrac{\mu}{2c_s}v$ is expected to decrease to zero in discrete steps corresponding to each wave reflection from the system's boundaries. Experimental evidence for the stepwise nature of the decrease in the radiation damping term (associated with discrete wave reflections) will be discussed below.

\subsection{A perturbative approach for rapid rupture and the slow rupture limit}
\label{subsec:perturbative}

The main theoretical prediction in Eqs.~\eqref{eq:v_r}-\eqref{eq:SD} has been obtained under the assumption that the spatiotemporal integral term $s(x,t)$ in Eq.~\eqref{eq:RD} vanishes well behind the propagating rupture.
While this idealized assumption is physically sensible, and is reasonably supported by simulational and experimental results for rapid rupture to be discussed below, the long-range spatiotemporal nature of $s(x,t)$ may suggest that it does not strictly vanish in many realistic situations. Consequently, our goal here is to understand how $v^0_{\rm res}$ and $\Delta\tau$ of Eqs.~\eqref{eq:v_r}-\eqref{eq:SD} change in the presence of a finite, yet small, $s(x,t)$.

To address this question, we denote the typical value of $s(x,t)$ at the tail of a rapidly propagating rupture fronts by $s$ and consider a perturbed solution of the form $v\=v^0_{\rm res}+\delta{v}_{\rm res}$. We then expand Eq.~\eqref{eq:RD}, which in the present context takes the form $\Delta\tau(v)=\frac{\mu}{2c_s}v-s$, to linear order in $\delta{v}_{\rm res}$ (it is already linear in $s$) to obtain
\begin{eqnarray}
\frac{\delta{v}_{\rm res}}{v^0_{\rm res}}=2\frac{c_s}{v^0_{\rm res}}\frac{s}{\mu}\left(1+\epsilon\right)^{-1}\ , \qquad\quad \epsilon \equiv 2\frac{c_s}{v^0_{\rm res}}\frac{\sigma}{\mu}\frac{\partial{f}_{\rm ss}}{\partial\log(v^0_{\rm res})}\nonumber\\
\Delta\tau + \delta(\Delta\tau)=\frac{\mu}{2 c_s}v^0_{\rm res}\left(1-\epsilon\,\frac{\delta{v}_{\rm res}}{v^0_{\rm res}}\right)\ .\hspace{2cm}
\label{eq:s_correction1}
\end{eqnarray}
Note that the internal state field $\phi$ has been assumed above to be slaved to $v$ and that $f_{\rm ss}(v)\=\tau_{\rm ss}(v)/\sigma$ is the steady-state friction coefficient.

Equation~\eqref{eq:s_correction1} reveals an interesting result; while both $v^0_{\rm res}$ and $\Delta\tau$ attain corrections that are linear in $s$, as expected from a linear perturbation approach, the actual smallness of the corresponding corrections may be quite different due to the appearance of $\epsilon$. $\epsilon$ is a product of $c_s/v^0_{\rm res}\!\gg\!1$, $\sigma/\mu\!\ll\!1$ and $\left|\frac{\partial{f}_{\rm ss}}{\partial\log(v^0_{\rm res})}\right|\!\ll\!1$~\cite{Marone1998a, Baumberger2006}, where the latter two contributions are expected to dominate the first one, leading to $\left|\epsilon\right|\!\ll\!1$. Consequently, while $\epsilon$ is expected to have a negligible effect on $\delta{v}_{\rm res}$, due to the appearance of the factor $(1+\epsilon)^{-1}$, it implies that $\delta(\Delta\tau)/\Delta\tau$ is a factor $\epsilon$ smaller than $\delta{v}_{\rm res}/v^0_{\rm res}$. We thus expect the stress drop $\Delta\tau$ to be far less sensitive to finite values of $s(x,t)$ compared to the residual slip velocity $v_{\rm res}$. This will be demonstrated below.

Equations~\eqref{eq:v_r}-\eqref{eq:s_correction1} have been derived under the assumption of vanishing or small spatiotemporal contribution $s(x,t)$, valid for physical situations in which rapid rupture emerges. Yet, when rupture velocities are negligible compared to elastic wave-speeds, i.e.~when slow rupture emerges~\cite{Peng2010,Obara2016,Takagi2016,Gomberg2016}, characteristic slip velocities $v$ are expected to be small such that the assumption behind Eqs.~\eqref{eq:v_r}-\eqref{eq:SD} may no longer be valid. In fact, for sufficiently slow rupture we expect the spatiotemporal integral term $s(x,t)$ in Eq.~\eqref{eq:RD} to be significantly {\em larger} than $\frac{\mu}{2c_s}v(x,t)$ such that
\begin{equation}
\frac{2c_s\Delta\tau}{\mu\,v_{\rm res}}\!\gg\!1\quad\hbox{and}\quad \Delta\tau\!\simeq\!-s(t)\qquad\hbox{for slow rupture} \ ,
\label{eq:slow}
\end{equation}
where $s(t)$ is determined by
\begin{equation}
s(t)\simeq\frac{\mu'}{2\pi}\!\int^\infty_{-\infty}\!\frac{\partial_{x'}\delta(x',t)}{x'-x}dx'\quad\hbox{\bf for sufficiently large}~~x \ .
\label{eq:QS}
\end{equation}
Here $\delta(x,t)$ is the slip displacement ($\dot\delta(x,t)\=v(x,t)$), and $\mu'\=\mu$ for antiplane shear and $\mu'\=\mu/(1-\nu)$ for in-plane shear ($\nu$ is Poisson's ratio). Equation~\eqref{eq:QS} is the quasi-static limit of the fully inertial integral term $s(x,t)$~\cite{Weertman1965}, which is expected to be valid for slow rupture, where inertial effects are negligible.

It is important to note that $s(t)$ of Eq.~\eqref{eq:QS}, and consequently also the stress drop $\Delta\tau$, may feature a non-trivial dependence on the rupture size and may attain finite values for long times in practical applications. In fact, a simple self-consistent dimensional analysis in the limit of large rupture in an infinite system indicates that $s$, and hence also $\Delta\tau$, inversely scales with the square root of the rupture size. Furthermore, note that while the stress drop $\Delta\tau$ predicted for slow rupture in Eq.~\eqref{eq:slow} is not related to the radiation damping term, it is still not a purely interfacial property, but rather it involves the long-range elasticity of the bodies surrounding the interface.

\subsection{Estimates of the spatiotemporal integral contribution to the stress drop}
\label{subsec:s_estimates}

In the previous subsection, we discussed the contribution of the long-range spatiotemporal term $s(x,t)$ to the stress drop, denoted there by $s$. In most of the discussion, excluding the last part concerning slow rupture, $s$ has been assumed to be small compared to the radiation damping contribution. However, we did not provide quantitative estimates for the relative magnitude of the two contributions, which was shown to be quantified by the ratio $\delta v_{\rm res}/v^0_{\rm res}$ (cf.~Eq.~\eqref{eq:s_correction1}). In this subsection, we provide such estimates in some strongly dynamic limiting cases. The basic idea is that since $s$ emerges from long-range linear elastodynamics, one can use some benchmark Linear Elastic Fracture Mechanics (LEFM) crack solutions~\cite{Broberg1999Book}, in which a finite stress drop $\Delta\tau$ is prescribed, to estimate its relative magnitude.

While the whole purpose of the present paper is to show that $\tau_{\rm res}$ is not an priori known interfacial quantity, and hence also $\Delta\tau\=\tau_{\rm d}-\tau_{\rm res}$ is not a priori known, LEFM solutions are still useful in estimating the relative magnitude of $s$ through $\delta v_{\rm res}/v^0_{\rm res}$. The point is that one can use the prescribed $\Delta\tau$ to obtain $v^0_{\rm res}$ according to Eq.~\eqref{eq:SD} for $s\=0$, then extract from the benchmark LEFM crack solution $v_{\rm res}$ for $s\!\ne\!0$ and finally obtain $\delta v_{\rm res}/v^0_{\rm res}\=(v_{\rm res}-v^0_{\rm res})/v^0_{\rm res}$ as an estimate for the relative magnitude of $s$ due to long-range linear elastodynamic interactions, which are properly captured by the crack solution.

To apply this procedure, we consider Broberg's self-similar crack solutions in infinite media~\cite{Broberg1999Book}. In these solutions, two crack tips are assumed to symmetrically expand at a constant velocity $c_{\rm r}$, starting from a zero crack length at $t\=0$. Self-similarity, which significantly simplifies the problem, implies that all fields in the problem depend only on the dimensionless combination $x/c_{\rm r} t$. While in general, as well as in the present work, frictional rupture propagation is not a self-similar process and the propagation speed is typically not constant, when the propagation speed approaches the limiting/asymptotic speed, self-similar conditions are reasonably approximated. Self-similar solutions have also been used in a related analysis in~\cite{Noda2009} (cf.~Appendix C3 therein), with the aim of deriving precise conditions for rupture mode selection (self-healing pulses vs.~crack-like), originally discussed in~\cite{Zheng1998}.

Applying these ideas to anti-plane shear (mode-III symmetry) self-similar crack solutions (see Eq.~(6.9.145) in~\cite{Broberg1999Book}), we obtain in the limit $c_{\rm r}\!\to\!c_s$ ($c_s$ is the limiting speed for mode-III), $\delta v_{\rm res}/v^0_{\rm res}\=2/\pi-1\!\approx\!-0.36$ (see Appendix A7). This result, which is in fact quantitatively supported by numerical simulations in Sect.~\ref{sec:simulations}, suggests that while the relative magnitude of $s$ is not extremely small under strongly dynamic mode-III conditions, it is still reasonably well within the range of validity of the linear perturbation theory of Eq.~\eqref{eq:s_correction1}. Repeating the same procedure for in-plane shear (mode-II symmetry) self-similar crack solutions (see Eq.~(6.9.85) in~\cite{Broberg1999Book}) in the limit $c_{\rm r}\!\to\!c_R$ ($c_R$ is the Rayleigh wave-speed, the limiting speed for mode-II), we obtain that $\delta v_{\rm res}/v^0_{\rm res}$ varies between $-0.613$ and $-0.417$ (the exact value depends on Poisson's ratio), see Appendix A7 for details.

This latter estimates are quantitatively similar to the mode-III estimate in the corresponding limiting case, demonstrating that $|s|$ is smaller than the radiation damping contribution for in-plane frictional rupture under strongly dynamic conditions. In both cases, the linear perturbation theory of Eq.~\eqref{eq:s_correction1} seems to be reasonably valid. We note that while the estimates above were based on bilateral (self-similar) rupture propagation, similar results are expected to emerge for unilateral propagation. Finally, we would like to stress again that the influence of $s$ on $\Delta\tau$ through $\delta v_{\rm res}/v^0_{\rm res}$ in Eq.~\eqref{eq:s_correction1} is strongly reduced by the small dimensionless factor $\epsilon$ (defined there). This discussion concludes the presentation of our main theoretical predictions, encapsulated in Eqs.~\eqref{eq:v_r}-\eqref{eq:QS}. In the next sections, we provide simulational and experimental support to these predictions.

\section{Simulational support}
\label{sec:simulations}

At this point, we first set out to test the predictions in Eqs.~\eqref{eq:v_r}-\eqref{eq:QS} against extensive numerical simulations. To that aim, we consider two semi-infinite bodies in frictional contact. The advantage of considering infinite-height bodies, i.e.~the $H\!\to\!\infty$ limit, is that the interfacial relation in Eq.~\eqref{eq:RD} becomes exact at all times, unlike for finite bodies. We also employ periodic boundary conditions, with periodicity $W$, in the sliding direction. We performed spectral boundary integral method~\cite{Geubelle1995,Morrissey1997,Breitenfeld1998} calculations under anti-plane shear (mode-III symmetry) deformation conditions, which are similar to --- yet somewhat simpler than --- the in-plane shear (mode-II symmetry) deformation conditions considered up to now~\cite{Freund1998}. The main simplification is that the displacement field in the mode-III problem ${\bm u}(x,y,t)\=u_z(x,y,t)\hat{\bm z}$ (the unit vectors satisfy $\hat{\bm z}\,\bot\,\hat{\bm x},\hat{\bm y}$) is essentially scalar. The basic field $u_z(x,y,t)$ satisfies the bulk elastodynamic equation $\mu\nabla^2 u_z\=\rho\,\ddot{u}_z$, together with $v(x,t)\!\equiv\!\dot{u}_z(x,y\=0^+,t)\!-\!\dot{u}_z(x,y\=0^-,t)$ and $\tau(x,t)\!\equiv\!\sigma_{yz}(x,y\=0,t)\=\mu\,\partial_yu_z(x,y\=0,t)$. Equation~\eqref{eq:RD} remains valid, where $\sigma_{xy}(x,y\=0,t)$ is replaced by $\sigma_{yz}(x,y\=0,t)$ and the integral term $s(x,t)$ corresponds to mode-III, see Appendices for more details.

The employed interfacial constitutive law features the generic properties discussed above, with $f(|v|,\phi)$ of Eq.~\eqref{eq:friction_law} that reduces under steady-state conditions to either the $N$-shaped or the no-minimum curves of Fig.~\ref{fig:Fig2}a (the exact expressions for $f(|v|,\phi)$ can be found in Appendix A2), and with $g(\cdot)$ of Eq.~\eqref{eq:dot_phi} that is given by $g\=1-|v|\phi/D$~\cite{Ruina1983,Baumberger2006,Bhattacharya2014,Marone1998a,Nakatani2001}. The bodies are loaded by a constant driving stress $\tau_{\rm d}$, as depicted schematically in Fig.~\ref{fig:Fig2}a, and frictional rupture is nucleated by introducing Gaussian perturbations of proper amplitude into a homogeneous state of very low slip velocity $v_0$ (that corresponds to the leftmost intersection point in Fig.~\ref{fig:Fig2}a), following the theoretical framework of~\cite{Brener2018} (the details of the nucleation procedure are described in the Appendices).

\begin{figure}[ht!]
  \centering
  \includegraphics[width=\columnwidth]{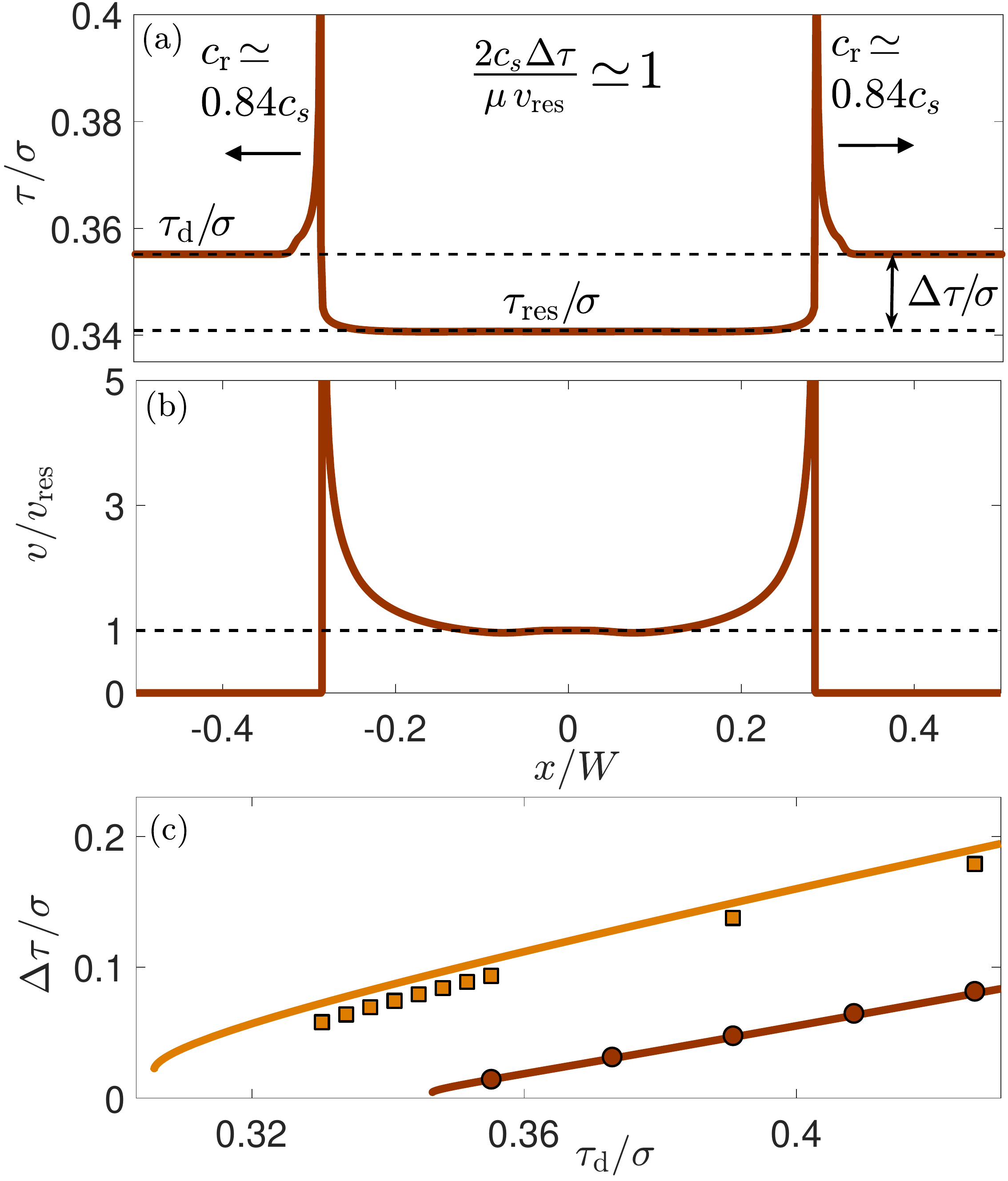}
  \caption{(a) A snapshot of the frictional stress $\tau(x)$ (normalized by $\sigma$) during rupture propagation that emerges in dynamic simulations with the $N$-shaped steady-state friction law of Fig.~\ref{fig:Fig2}a and $\tau_{\rm d}\!=\!0.355\sigma$ (see text for details). The snapshot reveals two rapid rupture fronts propagating at an instantaneous speed $c_{\rm r}\!\simeq\!0.84c_s$ in opposite directions into regions characterized by the applied stress $\tau_{\rm d}$ and leaving behind them a well-defined residual stress $\tau_{\rm res}\!<\!\tau_{\rm d}$. Consequently, a well-defined and finite stress drop $\Delta\tau$ emerges. Note that the $y$-axis is truncated at $\tau/\sigma\!=\!0.4$ for visual clarity (the maximal value of $\tau/\sigma$ is $0.58$) and that $x$ is normalized by the system length $W$ (the $x$-axis is shared with panel (b)). (b) The slip velocity $v(x)$ that corresponds to the snapshot shown in panel (a), normalized by the simulationally-measured residual velocity $v_{\rm res}$ (see text for discussion). The $y$-axis is also truncated for visual clarify and $x$ is normalized by the system length $W$. (c) The theoretical predictions of Eqs.~\eqref{eq:v_r}-\eqref{eq:SD} for $\Delta\tau(\tau_{\rm d})$ of rapid rupture (solid lines), both for the $N$-shaped steady-state friction law of Fig.~\ref{fig:Fig2}a (solid brown line, lower curve) and for the no-minimum law of Fig.~\ref{fig:Fig2}a (solid orange line, upper curve). As expected (cf.~Figs.~\ref{fig:Fig2}b-c), the former is smaller than the latter. The corresponding numerical results, obtained from the spatial stress distribution of frictional rupture such as the one shown in panel (a), are shown by the discrete symbols (circles for the $N$-shaped law, where the leftmost data point corresponds to the results shown in panel (a), and squares for the no-minimum law). See text for additional discussion.}\label{fig:Fig3}
\end{figure}

An example of the stress distribution of an emerging frictional rupture is shown in Fig.~\ref{fig:Fig3}a. The figure reveals two rapid rupture fronts propagating in opposite directions (at $84\%$ of the shear wave-speed $c_s$), where the stress ahead of the two fronts is the applied stress $\tau_{\rm d}$. As the emerging rupture is rapid, i.e.~propagating at a speed comparable to the elastic wave-speed, the relevant prediction is given by  Eq.~\eqref{eq:SD}. As predicted, the observed stress left behind the two rapid fronts, $\tau_{\rm res}$, is constant and smaller than the driving stress $\tau_{\rm d}$, giving rise to a finite stress drop $\Delta\tau$. Following the discussion above, since in these calculations $H\!\to\!\infty$, the finite stress drop persists indefinitely (while in finite size systems it persists for times $\sim{\cal O}(H/c_s)$, cf.~the experiments of~\cite{Rubino2017}, to be discussed later). The stress drop $\Delta\tau$ observed in Fig.~\ref{fig:Fig3}a quantitatively agrees with the prediction in Eq.~\eqref{eq:SD}, as stated in the figure legend.

In Fig.~\ref{fig:Fig3}b we present the slip velocity distribution that corresponds to the snapshot shown in Fig.~\ref{fig:Fig3}a. As predicted, the slip velocity attains a plateau level $v_{\rm res}$ behind the propagating rupture fronts, and the residual velocity $v_{\rm res}$ is used to normalize the slip velocity distribution. The simulationally-measured residual velocity $v_{\rm res}$ deviates by $\sim\!35\%$ from the theoretical prediction  in Eq.~\eqref{eq:SD}, in quantitative agreement with the estimate provided in Sect.~\ref{subsec:s_estimates} for the  spatiotemporal integral contribution to the stress drop in this case (see also Appendix A7). Consequently, while the upper equation in~\eqref{eq:s_correction1} implies that $s$ is in fact not very small, the prediction for $\Delta\tau$ works perfectly fine due to the $\epsilon\!\ll\!1$ factor in the lower equation in~\eqref{eq:s_correction1}.

In order to quantitatively test the latter prediction for rapid rupture over a range of physical conditions, we first solved Eq.~\eqref{eq:v_r} to obtain $v^0_{\rm res}(\tau_{\rm d})$ and then plugged it in Eq.~\eqref{eq:SD} to obtain $\Delta\tau(\tau_{\rm d})$, where the latter is plotted in solid lines in Fig.~\ref{fig:Fig3}c (the two solid lines correspond to both the $N$-shaped or the no-minimum steady-state friction laws shown in Fig.~\ref{fig:Fig2}a). We then numerically calculated $\Delta\tau$ (as demonstrated in Fig.~\ref{fig:Fig3}a) for various driving stresses $\tau_{\rm d}$, for both the $N$-shaped or the no-minimum steady-state friction curves shown in Fig.~\ref{fig:Fig2}a. The numerical results (discrete symbols) are superimposed on the theoretical prediction in Fig.~\ref{fig:Fig3}c (the lowest numerical data point on the lower curve corresponds to panels (a)-(b)). The agreement between the theoretical prediction and the numerical results for $\Delta\tau$ is very good, where it is better for the $N$-shaped steady-state friction law (lower curve) than for the no-minimum steady-state friction law (upper curve). This difference is fully accounted for by the magnitude and sign of $\epsilon$, directly affected by the $\frac{\partial{f}_{\rm ss}}{\partial\log(v^0_{\rm res})}$ term (cf.~Eq.~\eqref{eq:s_correction1}).

\begin{figure}[ht!]
  \centering
  \includegraphics[width=\columnwidth]{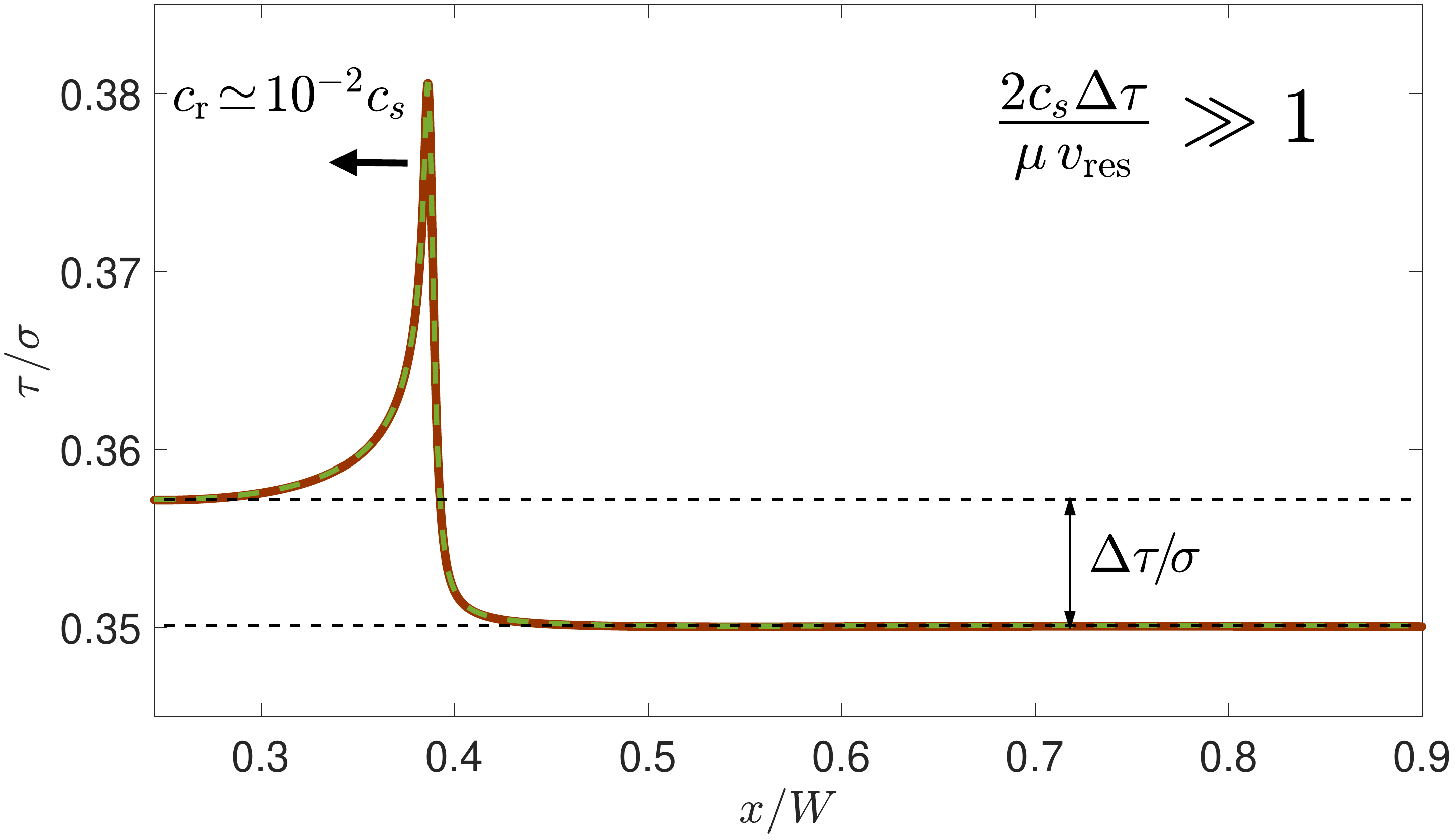}
  \caption{A snapshot of the frictional stress $\tau(x)$ (normalized by $\sigma$, $x$ is normalized by the system length $W$) of a slow rupture (solid line) propagating at about $1\%$ of the shear wave-speed $c_s$ and leaving behind it a stress drop $\Delta\tau$ (see Appendix A5 for details. Note that the other rupture edge is not shown). The stress drop satisfies $\frac{2c_s\Delta\tau}{\mu\,v_{\rm res}}\!\gg\!1$, as stated in the legend and in agreement with Eq.~\eqref{eq:slow}. $\tau_{\rm d}+s(x,t)$ (dashed green line), where $s(x,t)$ is given by the quasi-static integral of Eq.~\eqref{eq:QS}, is superimposed. See text for additional discussion.}
  \label{fig:Fig4}
\end{figure}

The contribution of the spatiotemporal integral $s(x,t)$, which is smaller than the radiation damping contribution for rapid rupture, may be a dominant effect for slow rupture, as discussed around Eqs.~\eqref{eq:slow}-\eqref{eq:QS}. To test this possibility, we generated slow rupture by changing the frictional parameters, the loading level and the nucleation procedure, as explained in Appendix A5. An example of such slow rupture is shown in Fig.~\ref{fig:Fig4} (solid line), exhibiting a rupture front propagating at a velocity $2$ orders of magnitude smaller than $c_s$ and leaving behind it a stress drop $\Delta\tau$. We first verified that $\frac{2c_s\Delta\tau}{\mu\,v_{\rm res}}\!\gg\!1$, as predicted by Eq.~\eqref{eq:slow} (and as stated in the figure legend). To test whether indeed $\Delta\tau\!\simeq\!-s(t)$, cf.~Eq.~\eqref{eq:slow}, we used the slip displacement gradient $\partial_{x}\delta(x,t)$ obtained from the dynamical simulation and used it to calculate the quasi-static integral in Eq.~\eqref{eq:QS} for any point $x$ along the interface at $t$ corresponding to the snapshot in Fig.~\ref{fig:Fig4}. The result for $\tau_{\rm d}+s(x,t)$ is superimposed on $\tau(x,t)$ of Fig.~\ref{fig:Fig4} (dashed line), demonstrating excellent agreement with the fully dynamic result, and in particular validating $\Delta\tau\!\simeq\!-s(t)$ for this $t$. Consequently, our simulations strongly support the theoretical predictions in Eqs.~\eqref{eq:v_r}-\eqref{eq:QS}, for both rapid and slow frictional rupture. Next, we discuss direct experimental support for these predictions.

\section{Experimental support}
\label{sec:exp}

In the previous section, we provided strong simulational support to the theoretical predictions in Eqs.~\eqref{eq:v_r}-\eqref{eq:QS}. Our goal here is to test these predictions against experimental data. Equation~\eqref{eq:v_r} predicts the slip velocity $v_{\rm res}$ behind the frictional rupture once the steady-state friction curve $\tau_{\rm ss}(v)$ is known. The latter is not always known a priori over a sufficiently wide range of steady-state slip velocities. In fact, measuring both the frictional stress $\tau_{\rm res}$ and the slip velocity $v_{\rm res}$ behind rupture fronts allows to extract $\tau_{\rm ss}(v)$. In this case, any triplet $(\tau_{\rm d}, \tau_{\rm res}, v_{\rm res})$ is predicted to follow either Eq.~\eqref{eq:SD} for rapid rupture or Eq.~\eqref{eq:slow} for slow rupture, {\em independently} of the steady-state friction law $\tau_{\rm ss}(v)$.

Measurements of both $\tau_{\rm res}$ and $v_{\rm res}$ behind rupture fronts for various $\tau_{\rm d}$ have been recently performed by two independent experimental groups using two different experimental systems and techniques~\cite{Rubino2017,Svetlizky2017a}.  The first focused on the frictional dynamics along the interface between two blocks of Homalite probed through a novel ultrahigh full-field imaging technique~\cite{Rubino2017}. The second focused on the frictional dynamics along the interface between two blocks of poly(methylmethacrylate) (PMMA) probed through a combination of high speed interfacial imaging (via a method of total internal reflection) and simultaneous measurements of the deformation fields slightly above the interface~\cite{Svetlizky2017a}. Here we use the data reported in these works to quantitatively test our predictions.

\begin{figure}[ht!]
  \centering
  \includegraphics[width=0.9\columnwidth]{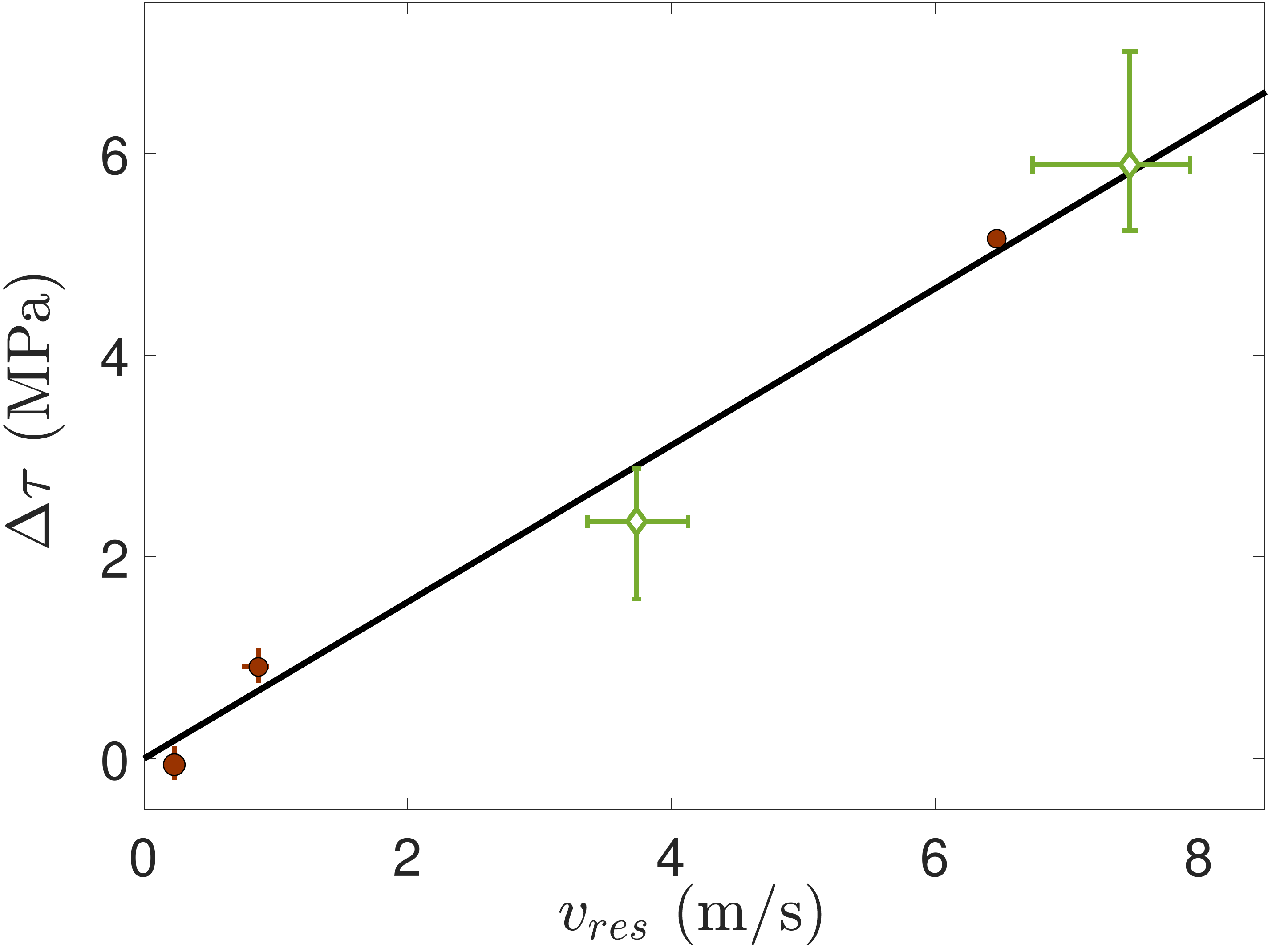}
  \caption{The theoretical prediction of Eq.~\eqref{eq:SD} is plotted for Homalite by the solid line, with the following high strain rates material parameters $\mu\!=\!1.96$GPa and $c_s\!=\!1263$m/s~\cite{Rubino2017,Singh2003}, which uniquely determine the slope of the stress drop $\Delta\tau$ vs.~the slip rate $v_{\rm res}$ behind a rupture front. Experiments on the rupture of frictional interfaces composed of two blocks of Homalite have been performed in~\cite{Rubino2017}, where a novel ultrahigh full-field imaging technique has been employed to directly measure $\tau_{\rm res}$ and $v_{\rm res}$ behind rupture fronts. Experiments for different values of the applied shear stress $\tau_{\rm d}$ have been performed, which allowed us to extract from Fig.~8a of~\cite{Rubino2017} five triplets $(\tau_{\rm d}, \tau_{\rm res}, v_{\rm res})$ (see details in Appendix A6). For each triplet we calculated $\Delta\tau\!\equiv\!\tau_{\rm d}-\tau_{\rm res}$, and then superimposed the resulting $\Delta\tau$ vs.~$v_{\rm res}$ on the theoretical prediction (discrete symbols). The symbols (and colors) differentiate data obtained from high-resolution (full brown circles) and low-resolution (empty green diamonds) measurements, following the classification of~\cite{Rubino2017}. The symbol size and/or error bar represent the full range of measurements reported on in Fig.~8a of~\cite{Rubino2017} per imposed far-field conditions.}
  \label{fig:Fig5}
\end{figure}

We start with~\cite{Rubino2017}, where the observed rupture fronts were all in the rapid (fully inertial) regime, and hence the relevant prediction is given in Eq.~\eqref{eq:SD}. To test this prediction, we extracted the relevant experimental data from~\cite{Rubino2017} and plotted in Fig.~\ref{fig:Fig5} the resulting $\Delta\tau$ vs.~$v_{\rm res}$ against the linear $\Delta\tau(v_{\rm res})$ relation of Eq.~\eqref{eq:SD}, with a slope that corresponds to $\mu/2c_s$ of Homalite (see figure caption for additional details). The results reveal excellent agreement between the experimental data and the theoretical prediction, without any free parameter. The theoretical relation in Eq.~\eqref{eq:SD} predicts $v^0_{\rm res}$ by assuming that the long-range spatiotemporal contribution $s(x,t)$ is negligibly small, while in the experimental data we just use the measured $v_{\rm res}$. The fact that there are no significant horizontal deviations of the experimental data from the theoretical line of Eq.~\eqref{eq:SD}, i.e.~that $v_{\rm res}$ is close to $v^0_{\rm res}$, indicates that $s(x,t)$ is smaller in these experiments compared to the estimate provided in Sect.~\ref{subsec:s_estimates}. The origin of the latter discrepancy is not yet clear.

The experimental data included in Fig.~\ref{fig:Fig5} have been obtained in the short time regime, before any wave reflection from the system's boundaries, for which Eqs.~\eqref{eq:v_r}-\eqref{eq:SD} are valid. Yet, for a single case, measurements are reported {\em after} the first wave reflection (but before the second one), cf.~Fig.~4c in~\cite{Rubino2017}. Under these conditions, we expect the ordinary radiation damping term $\tfrac{\mu}{2c_s}v$ to be replaced by a {\em smaller} term. Consequently, Eq.~\eqref{eq:v_r} predicts that $v_{\rm res}$ should {\em increase}, exactly as is experimentally observed in Fig.~4c in~\cite{Rubino2017} (denoted as ``Reflected rupture'' in the figure), thus providing direct evidence for the stepwise reduction of the radiation damping term with discrete wave reflections.

We now turn to the experiments of~\cite{Svetlizky2017a}, which report on extensive measurements of $\tau_{\rm res}$ and $v_{\rm res}$, cf.~Fig.~3a there. In order to test our predictions we need also measurements of the driving stress $\tau_{\rm d}$, which will allow us to extract the stress drop $\Delta\tau\=\tau_{\rm d}-\tau_{\rm res}$. The corresponding extensive measurements of $\tau_{\rm d}$ are not presented in~\cite{Svetlizky2017a}, though $\tau_{\rm d}$ is presented for two highly relevant examples in Fig.~2b there (where $\tau_{\rm d}$ is denoted by $\tau_0$). One example corresponds to very rapid rupture (in fact, it is supershear rupture, propagating at nearly the dilatational wave-speed) and the other to slow rupture (propagating at about $10\%$ of the Rayleigh wave-speed). These data are exactly what is needed in order to test our predictions, in particular regarding the change in behavior for slow rupture. For rapid rupture, we extracted from Fig.~2b of~\cite{Svetlizky2017a} (blue data) $\Delta\tau\=1.17$ MPa and from Fig.~1d (bottom) $v_{\rm res}\!\simeq\!1.5$ m/s (see additional details in Appendix A6). Using the reported values $\rho\=1170$ kg/m$^3$ and $c_s\=1345$ m/s~\cite{Svetlizky2017a}, together with $\mu\=\rho c_s^2$, we obtain $\frac{2c_s\Delta\tau}{\mu\,v_{\rm res}}\=0.99$. This is in great quantitative agreement with the prediction for rapid rupture in Eq.~\eqref{eq:SD}, and is fully consistent with the independent experimental data presented in Fig.~\ref{fig:Fig5}. This result indicates that the long-range spatiotemporal contribution $s(x,t)$ is indeed small also in the experiments of~\cite{Svetlizky2017a}.

We then extracted the corresponding data for the slow rupture, obtaining from Fig.~2b of~\cite{Svetlizky2017a} (red data) $\Delta\tau\=0.242$ MPa and from Fig.~1d (top) $v_{\rm res}\!\simeq\!0.02$ m/s. Using these data we obtain $\frac{2c_s\Delta\tau}{\mu\,v_{\rm res}}\!\simeq\!15\!\gg\!1$, in agreement with the theoretical prediction for slow rupture in Eq.~\eqref{eq:slow}. The results presented in the last two sections provide strong simulational and experimental support for our theoretical predictions and hence for the proposed picture of the physical origin of stress drops in frictional rupture. As stress drops are important for the possible emergence of crack-like behaviour in frictional rupture, future studies of the latter should be based on the results presented in this paper.

\section{Discussion and concluding remarks}
\label{sec:summary}

The possible deep relations between frictional rupture and ordinary fracture provide a powerful conceptual and quantitative framework to understand frictional dynamics in a wide variety of physical contexts. This framework is extensively used to interpret and quantify geophysical observations~\cite{Abercrombie2005,Bizzarri2016}, as well as a broad spectrum of laboratory phenomena~\cite{Lu2010,Lu2010a,Noda2013a,Svetlizky2014,Bayart2015,Svetlizky2016,Rubino2017}. For example, a recent series of careful laboratory experiments~\cite{Svetlizky2014,Bayart2015,Svetlizky2016} demonstrated that when the analogy between frictional rupture and ordinary fracture holds, the dynamic propagation of laboratory earthquakes and their arrest can be quantitatively understood to an unprecedented degree~\cite{Kammer2015}. Yet, the fundamental physical origin and range of validity of the analogy between frictional rupture and ordinary fracture are not yet fully understood. In this paper, we developed a comprehensive and fundamental understanding of why, how and to what extent frictional rupture might be viewed as an ordinary fracture process.

A important ingredient in the analogy is the emergence of a finite and well-defined stress drop $\Delta\tau\=\tau_{\rm d}-\tau_{\rm res}$, the difference between the applied driving stress $\tau_{\rm d}$ and the residual stress $\tau_{\rm res}$, in frictional rupture. In the first part of the paper we showed that, contrary to widely adopted assumptions, the residual stress $\tau_{\rm res}$ is not a characteristic property of frictional interfaces. Rather, for rapid rupture $\tau_{\rm res}$ is shown to crucially depend on elastodynamic bulk effects --- mainly wave radiation from the frictional interface to the bodies surrounding it, but also long-range elastodynamic bulk interactions (encapsulated in the integral term $s(x,t)$ in Eq.~\eqref{eq:RD})   --- and the applied driving stress $\tau_{\rm d}$ itself, in addition to the contribution of the slip rate dependence of the constitutive friction law. Notably, we showed that for rapid rupture the deviation of $\tau_{\rm res}$ from $\tau_{\rm d}$, i.e.~the existence of a finite stress drop $\Delta\tau$, is a finite time effect, mainly limited by the wave travel time in finite systems. For slow rupture, it is shown that if a stress drop exists, it is intimately related to the long-range quasi-static elasticity of the bodies surrounding the interface, again not exclusively to interfacial physics. Our theoretical predictions are supported by extensive computations and existing experimental data from two independent laboratory experiments.

Our findings have important implications that go beyond their basic nature; first, the results show that the widely used slip-weakening models~\cite{Ida1972,Palmer1973}, in which the existence of a residual stress $\tau_{\rm res}\!<\!\tau_{\rm d}$ is a priori assumed (as a fixed interfacial property), should be employed with care. In particular, as $\tau_{\rm res}$ has been shown to depend on the externally applied stress $\tau_{\rm d}$, on the properties of the bodies surrounding the interface and on the rate-dependence of the frictional constitutive behavior, $\tau_{\rm res}$ cannot be assumed to be fixed. Rather, it should be self-consistently calculated from the coupled interface-bulk problem. Somewhat related conclusions in relation to slip-weakening models, based on measurements of evolving local friction during spontaneously developing laboratory earthquakes, have been drawn in~\cite{Rubino2017}. We also note that the dependence of $\tau_{\rm res}$ on the bulk constitutive relation and on long-range bulk-mediated interactions may give rise to interesting effects for more complicated bulk constitutive relations (e.g.~viscoelasticity) and in the presence of repeated rupture events~\cite{Radiguet2013}. Such effects should be explored in future work.

The existence of a finite stress drop $\Delta\tau$ generically leads to accelerating frictional rupture under stress-controlled far-field loading conditions $\tau_{\rm d}$ if $\Delta\tau$ is independent of the rupture size $L$. In these situations, inertia-limited rapid rupture is expected to emerge on time scales for which wave interaction with the outer boundaries does not exist, and $\Delta\tau$ is controlled by elastodynamic bulk effects. These conditions are typically realized in many geological and laboratory earthquakes~\cite{Abercrombie2005,Lu2010,Lu2010a,Noda2013a,Svetlizky2014,Bayart2015,Bizzarri2016,Svetlizky2016,Rubino2017}. On the other hand, slow rupture propagation --- a widely observed, yet highly debated and elusive phenomenon~\cite{Peng2010,Obara2016,Takagi2016,Gomberg2016} --- is expected to feature smaller stress drops $\Delta\tau$ that may decrease with increasing rupture size $L$, such that rupture acceleration is limited.

The possible $L$ dependence of $\Delta\tau$ and its possible relations to the emergence of slow rupture should be further explored in the future. In addition, as the whole discussion in this paper is valid for stress-controlled far-field loading conditions (characterized by $\tau_{\rm d}$), future work should also consider velocity-controlled far-field loading conditions, where finite stress drops might emerge from different physical considerations~\cite{Brener2005}. Finally, as the existence of a finite stress drop $\Delta\tau$ does not in itself guarantee the crack-like behavior of frictional rupture, future work should clarify to what extent the analogy to ordinary cracks can in fact quantitatively account for the dynamics of frictional rupture. All in all, as stress drops are key quantities in frictional failure dynamics, we expect our results to provide a conceptual and quantitative framework to address various fundamental and applied problems in relation to the rupture dynamics of frictional interfaces, with implications for both laboratory and geophysical-scale phenomena.

{\em Acknowledgements} E.~B.~and J.-F.M.~acknowledge support from the Rothschild Caesarea Foundation. E.~B.~acknowledges support
from the Israel Science Foundation (Grant No.~295/16). J.-F.M., F.~B.~and T.~R.~acknowledge support from the Swiss
National Science Foundation (Grant No.~162569). This research is made possible in part by the historic generosity of the Harold Perlman Family. We are grateful to Ilya Svetlizky for pointing out the differences between rapid and slow rupture in the context of our work, and the relevance of the experimental data of~\cite{Svetlizky2017a} to it. His comment has stimulated the discussion of slow rupture in this paper. In addition, we are grateful to him for pushing us to look at Broberg's self-similar solutions, eventually leading to the extraction of the estimates presented in Sect.~\ref{subsec:s_estimates}.

\setcounter{equation}{0}
\setcounter{figure}{0}
\setcounter{section}{0}
\setcounter{table}{0}
\makeatletter
\renewcommand{\theequation}{A\arabic{equation}}
\renewcommand{\thefigure}{A\arabic{figure}}


\section*{Appendix A1:\\ The main numerical method}
\label{sec:num}

The simulations discussed above relied on a spectral boundary integral
formulation of the elastodynamic equations~\cite{Geubelle1995,Morrissey1997,Breitenfeld1998}. The
latter relates the traction stresses acting along the interface between two linearly
elastic half-spaces and the resulting displacements. For the mode-III (anti-plane shear) elastodynamic
problem studied in the manuscript, the interface is initially uniformly
pre-stressed by $\tau_{\rm d}$ and is set to slide at an extremely small steady velocity
$v_0$, such that the shear tractions at the interface take the form
\begin{equation}
\tau(x,t) = \tau_{\rm d} -\frac{\mu}{2c_s}\Big(v(x,t)-v_0\Big) + s(x,t) \ .
\label{spectral}
\end{equation}
The second right-hand-side term represents the instantaneous response to
changes in the sliding velocity, the so-called radiation damping term. As discussed in the
manuscript, this term can be understood as the damping of interfacial energy
due to elastic waves radiated into the infinite domain. The third term $s(x,t)$ accounts for the
history and spatial distribution of interfacial displacements $u(x,t)$. Both $s(x,t)$ and $u(x,t)$ are related in
the spectral domain via a convolution integral, whose expression can be found in
\cite{Breitenfeld1998}. Due to the spectral nature of the formulation, the
simulated domain is taken to be periodic in the lateral direction, with periodicity $W$. The latter is chosen to be large
enough to prevent any effect of the periodicity on the results reported in the manuscript.

Rupture is nucleated at the center of the domain by introducing a Gaussian perturbation of
the slip velocity into an initial steady sliding state at $v_0$.
The sliding velocity is then computed by combining Eq.~\eqref{spectral} and the rate and
state-dependent friction law $\tau=\sigma\mathrm{sgn}(v)f(|v|,\phi)$ (see Appendix A2 for
more details). $u(x,t)$ is then integrated in time using an explicit time-stepping scheme
\begin{eqnarray}
\label{eq:tstep1}
u(x,t+\Delta t) = u(x,t) + 0.5\,v(x,t)\,\Delta t \ .
\end{eqnarray}
Note that the factor $0.5$ on the right-hand-side of Eq.~\eqref{eq:tstep1} ensures that $v(x,t)$ is
indeed the slip velocity. In order to guarantee the stability and the convergence of the numerical
scheme, $\Delta t$ is defined as the time needed for a shear wave to travel a fraction $0.2$
of one grid spacing, i.e.~$\Delta t\=0.2\,\Delta x/c_s$. Additional information about the numerical
scheme and the nucleation procedure can be found in~\cite{Brener2018}, together with videos of similar rupture events.

\section*{Appendix A2: The friction laws}
\label{sec:fric}

The friction laws used in this work, and whose steady-state behaviors are plotted in Fig.~2, are related to the one used previously
in~\cite{Aldam2017a,Brener2018}. The friction law is defined by the relation between
the shear stress $\tau\!\equiv\!\sigma_{xy}$ and the compressive normal stress
$\sigma\!\equiv\!-\sigma_{yy}$ at the interface,  $\tau\=\sigma
\sgn(v)f\left(\left|v\right|,\phi\right)$, and by the evolution equation for state variable $\phi$,
$\dot\phi\=g\left(\left|v\right|,\phi\right)$. The constitutive functions
$f\left(\left|v\right|,\phi\right)$ and $g\left(\left|v\right|,\phi\right)$ used in this work take
the form
\begin{eqnarray}
\label{eq:f}
&& f\left(\left|v\right|,\phi\right)=\left[1+b \log \left(1+\frac{\phi }{\phi _*}\right)\right]
\times \\
&&\qquad\qquad\qquad\qquad \left[\frac{f_0}{\sqrt{1+\left(v_*/v\right)^2}}+a  \log
\left(1+\frac{\left| v\right| }{v_*}\right)\right]\ ,\nonumber\\
\label{eq:g}
&&g\left(\left|v\right|,\phi\right)=1-\frac{\left| v\right|\phi}{D}\sqrt{1+\left(v_*/v\right)^2}\ ,
\end{eqnarray}
where $\phi$ represents the typical age/maturity of contact asperities that compose the
interface at a microscopic scale~\cite{Baumberger2006}. In Eq.~\eqref{eq:f}, $f_0$
sets the scale of the dimensionless frictional resistance (friction coefficient), $b$ is the aging
coefficient and $a$ is related to the thermally-activated rheology of
contact asperities~\cite{Baumberger2006}. The function $\sqrt{1+\left(v_*/v\right)^2}$
that appears also in $g\left(\left|v\right|,\phi\right)$ ensures that for vanishingly small
steady-state velocities, $\phi$ saturates after extremely long times to a finite value of $D/v_*$,
rather than diverges. As discussed in \cite{Brener2018}, this regularization makes no
significant difference in the results discussed above. For the sake of
notational simplicity, the regularization is hence omitted in the main text (though it is included in
the calculations). Eqs.~\eqref{eq:f} and ~\eqref{eq:g} lead to the steady-state friction curve of Fig.~2 with a minimum at an intermediate $v$ (brown solid line), while
the no-minimum steady-state friction curve (dash-dotted orange line) is obtained after
neglecting the ``+1'' in the $b$ term.

The reader is referred to \cite{Aldam2017a,Brener2018} for additional discussions about
the formulations of Eqs.~\eqref{eq:f}-\eqref{eq:g}, which go beyond the conventional
rate-and-state friction laws. Nevertheless, the results and conclusions discussed above
are independent of the choice of the rate-and-state formulation.

\section*{Appendix A3: 1D rupture fronts}

In this Appendix we describe propagating steady-state rupture fronts in thin (quasi-1D) systems, where no stress drops emerge. We consider two
long and thin linear elastic bodies of height $H$ in frictional contact, such that the momentum balance equation $\rho\ddot{\bm
u}\=\nabla\!\cdot{\bm \sigma}$ reduces to~\cite{Bar-Sinai2012,Bar-Sinai2013,Brener2018}
\begin{equation}
\label{eq:FB}
H\bar\mu\left(c_{\rm 1D}^{-2}\partial_{tt}-\partial_{xx}\right)\!u\!\left(x,t\right)\!=\!\tau_{\rm
d}\!-\tau\left[v\!\left(x,t\right),\phi\!\left(x,t\right)\right]\ ,
\end{equation}
where $u\!\equiv\!u_x$, $\bar\mu$ and $c_{\rm 1D}$ are the effective shear modulus and
wave-speed~\cite{Bar-Sinai2012,Bar-Sinai2013}, respectively, and $\tau_{\rm d}$ is a constant
driving stress (see Fig.~2).

Propagating 1D steady-state solutions then satisfy~\cite{Brener2018}
\begin{eqnarray}
\label{eq:SS_FB}
&\bar\mu H c_{\rm 1D}^{-1}(1-\beta^2)\beta^{-1}\,v'(\xi)\!=\!\tau_{\rm d}\!-\!\tau (v(\xi ),\phi
(\xi ))\ ,\\
\label{eq:SS_phi}
&\beta\,c_{\rm 1D}\,\phi'(\xi)=\phi(\xi)v(\xi)/D-1\ ,
\end{eqnarray}
where we defined a co-moving coordinate $\xi\!\equiv\!x\!-\!\beta c_{\rm 1D}\,t$, integrated out
$u$ and eliminated partial time-derivatives.

\begin{figure}[ht]
\centering
\includegraphics[width=\columnwidth]{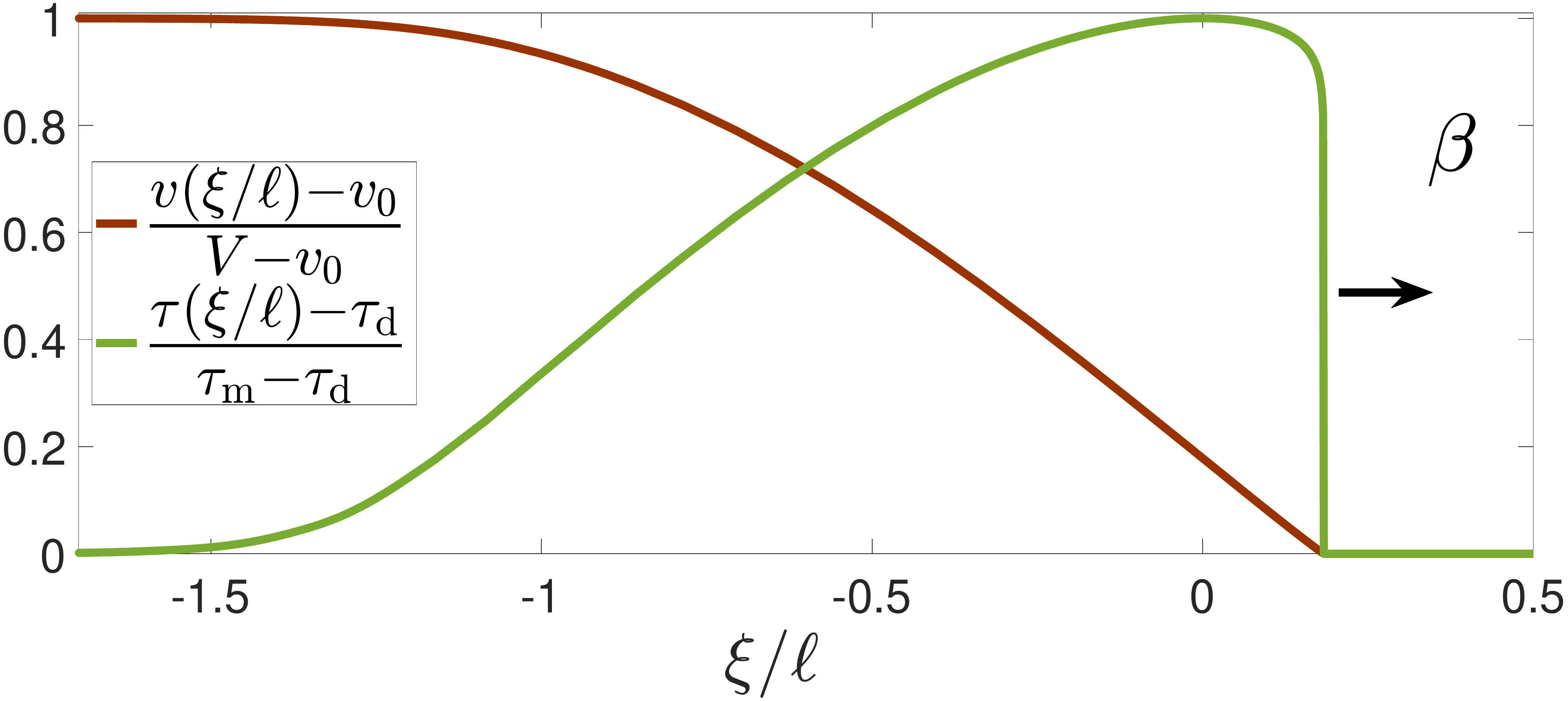}
\caption{The normalized spatial profiles of $\tau(\xi)$ and $v(\xi)$ near a steady-state rupture
front edge propagating from left to right with a velocity $c_{\rm r}^{\rm 1D}\!=\!\beta c_s$, with $\beta\!=\!0.144$ (see text for details on the employed
normalization). Note also that $\tau_{\rm d}/\sigma\!=\!0.355$, exactly as in Fig.~2, though in the latter 2D case a rupture front with $\beta\!=\!0.84$ emerged (cf.~Fig.~3a in the manuscript).}\label{fig:1Drup}
\end{figure}

Steady-state rupture propagation is a dynamical process in which a homogeneous $V$ state invades a homogeneous $v_0\!\ll\!V$
state~\cite{Bar-Sinai2012,Svetlizky2014,Rubino2017,Brener2018}, both shown in Fig.~2 as the intersections of the velocity
strengthening branches of the friction law with the driving stress $\tau_{\rm d}$. We found these
solutions for the friction law described in Sect.~\ref{sec:fric}, using a shooting
method~\cite{Press2007} (similar to that used in~\cite{Bar-Sinai2012,Bar-Sinai2013}). The solution is shown in
Fig.~\ref{fig:1Drup}. To normalize the stress fields we used the definition $\tau_{\rm m}$ is the maximal value $\tau$
attains in the profile, and $\ell\!\equiv\!\frac{1-\beta ^2}{\beta}\frac{H V \bar\mu}{c_{\rm 1D}
\left(\tau_{\rm m}-\tau_{\rm d}\right)}$ is the lengthscale over which the fields change, which can be
calculated by a scaling analysis of Eq.~\eqref{eq:SS_FB}.

As seen in Fig.~\ref{fig:1Drup}, the stress both ahead and behind the rupture front equal
$\tau_{\rm d}$, i.e.~there exists no stress drop. A corollary is that no singularity is observed in Fig.~\ref{fig:1Drup} (compare to Figs.~3a and 4).

\section*{Appendix A4: Parameters}
\label{sec:param}

The parameters used for all the calculations described in this work, except for the slow rupture to be discussed in Appendix A5, are given in Table~\ref{tab:values}.
\begin{table}[ht]
  \centering
  \begin{tabular}{|c|c|c|}
  \hline
  Parameter & Value & Units\\
  \hline
  $\mu$ , $\bar\mu$ & $9\!\times\!10^9$ & Pa\\ \hline
  $\sigma$ & $10^6$ & Pa\\ \hline
  $c_s$ , $c_{\rm 1D}$ & $2739$ & m/s\\ \hline
  $D$ & $5\!\times\!10^{-7}$ &m \\ \hline
  $b$ & $0.075$ & -\\ \hline
  $v_*$ & $10^{-7}$ & m/s\\ \hline
  $f_0$ & $0.28$ & -\\ \hline
  $\phi_*$ & $3.3\!\times\!10^{-4}$ & s\\ \hline
  $a$ & $0.005$ & -\\ \hline
  $h$ & $0.2$ &mm \\  \hline
\end{tabular}
  \caption{Values for all parameters used (in MKS units).}\label{tab:values}
\end{table}
Note that the values of the listed parameters are characteristic of some laboratory experiments (see~\cite{Bar-Sinai2014} for details). However, the generic properties of the derived results are independent of the exact numbers, and are relevant to a broad range of materials and physical situations. For example, $v_*$ that controls the velocity scale below which the system is in the stick phase, can be taken to be significantly smaller.

\section*{Appendix A5: 2D slow rupture fronts}
\label{sec:slow}

\begin{figure}
\centering
\includegraphics[width=\linewidth]{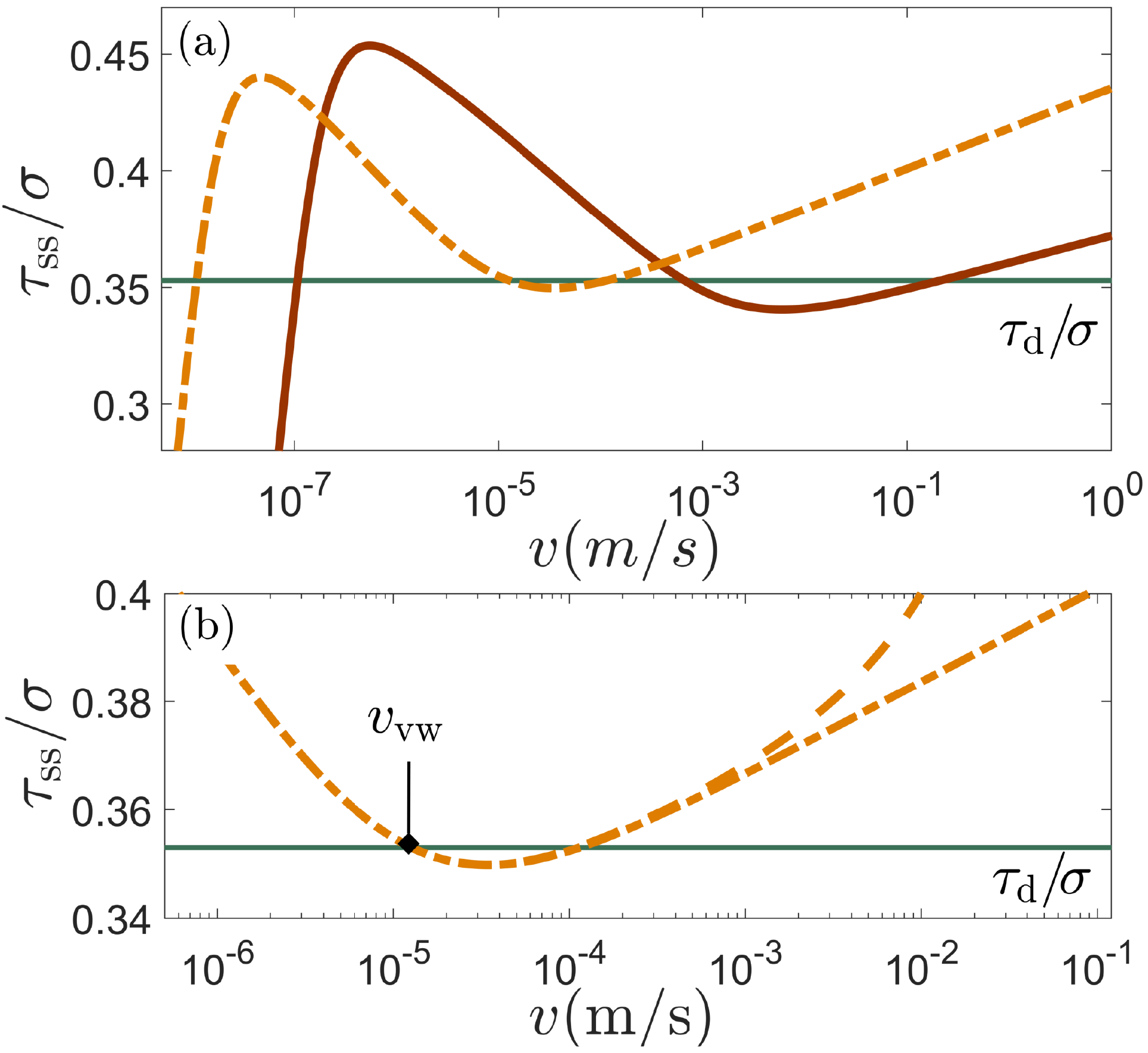}
\caption{(a) The normalized steady-state friction law as used in Fig.~2a (solid brown line) and the modified one that corresponds to the parameters listed in Table~\ref{tab:slowf_values} (dash-dotted orange line). The horizontal green line represents the normalized driving stress $\tau_{\rm d}/\sigma$. (b) A zoom-in on the dash-dotted orange line of panel (a), where the effective steady-state friction curve (dashed orange line, obtained by adding the radiation damping term $\frac{\mu}{2c_s}v$) is added. The intersection of the driving stress with the velocity-weakening branch of the friction law is denoted by $v_{\rm vw}$ (black diamond). Perturbations around $v_{\rm vw}$ leads to the slow rupture shown in Fig.~4, see text for additional details.}
\label{fig:slow_front_law}
\end{figure}

The very same constitutive framework can give rise to slow rupture fronts ($c_{\rm r}\!\ll\!c_s$), as demonstrated in Fig.~4. While the emergence of slow rupture is of great interest in general, in the present context we are just interested in generating slow rupture and studying its properties in relation to the theoretical prediction in Eqs.~(7)-(8). One way to generate slow rupture within our constitutive framework is to use friction parameters that shift the steady-state friction curve to smaller slip velocities and to employ a different nucleation procedure.

In particular, using the different set of parameters listed in Table~\ref{tab:slowf_values}, we obtain the dash-dotted orange steady-state curve in Fig.~\ref{fig:slow_front_law}a (the solid brown line is identical to the one shown in Fig.~2a). For these parameters, and for the same value of the normalized driving stress $\tau_{\rm d}/\sigma$, the effective steady-state friction curve shown in Fig.~\ref{fig:slow_front_law}b (dashed orange line, obtained by adding the radiation damping term $\frac{\mu}{2c_s}v$), is practically indistinguishable from the steady-state friction curve in the relevant slip velocities range. In addition, rupture is nucleated by introducing a perturbation to the internal state field $\phi$ of the form
\begin{equation}
\phi(x,t=0)=\frac{D}{v_{\rm vw}}+\varepsilon\sin(kx) \ ,
\label{equ:lin_instab}
\end{equation}
with $k\=2\pi/W$ and $\varepsilon\=10^{-4}$, into an interface that slides homogeneously at a velocity $v_{\rm vw}$ that corresponds to a fixed-point on the velocity-weakening branch (it is marked by the black diamond in Fig.~\ref{fig:slow_front_law}b). This nucleation procedure is different from the one used elsewhere in the paper, where Gaussian perturbations are introduced into an essentially locked-in interface, as described in detail in~\cite{Brener2018}.
\begin{table}[ht]
  \centering
  \begin{tabular}{|c|c|c|}
  \hline
  Parameter & Value & Units\\
  \hline
  $D$ & $5\!\times\!10^{-7}$ &m \\ \hline
  $b$ & $0.1$ & -\\ \hline
  $v_*$ & $10^{-8}$ & m/s\\ \hline
  $f_0$ & $0.28$ & -\\ \hline
  $\phi_*$ & $0.05$ & s\\ \hline
  $a$ & $0.0075$ & -\\
   \hline
\end{tabular}
  \caption{The values of the rate-and-state parameters (in MKS units), which are discussed in Fig.~\ref{fig:slow_front_law} and which gave rise to the slow rupture shown in Fig.~4.}\label{tab:slowf_values}
\end{table}

These modifications are sufficient to generate the slow rupture shown in Fig.~4. The physics behind the emergence of slow rupture, which is very interesting in itself, is not thoroughly discussed here. It deserves an investigation of its own, which we hope to pursue in the future.

\section*{Appendix A6: The experimental data}
\label{sec:expe}

The data reported in Fig.~5 are obtained from the experimental measurements
\cite{Rubino2017} of ruptures propagating along a frictional interface formed by two plates of Homalite. Figure~8 of~\cite{Rubino2017} reports the steady-state
friction coefficient versus slip rate measured at the interface in the wake of the propagating
rupture front. In terms of the notation used in this manuscript, the former is the ratio $\tau_{\rm
res}/\sigma$, while the latter corresponds to $v_{\rm res}$. In the experiments of~\cite{Rubino2017}, the frictional
interface is pre-cut at an angle $\alpha$ from the principal direction of the imposed compressive
stress $P$, such that
\begin{eqnarray}
  &\tau_{\rm d} = P\cos\alpha\sin\alpha, \\
  &\sigma = P\cos^2\alpha.
\end{eqnarray}

Rubino et al.~\cite{Rubino2017} distinguished data measured with respectively high and low levels of accuracy. From the data sets reported in Fig.~8 of~\cite{Rubino2017} and their associated boundary conditions listed in Table~\ref{tab:rubino}, we compute the triplets
$(\tau_{\rm d},\tau_{\rm res},v_{\rm res})$, which are then used to construct Fig.~5.
\begin{table}[ht]
  \centering
  \begin{tabular}{|c|c|c|c|}
  \hline
  Symbols & Resolution & $P$ [MPa] & $\alpha$ \\
  \hline
  Blue dots & High & $23$ & $29^{\circ}$\\  \hline
  Red dots & High & $7.4$ & $29^{\circ}$\\  \hline
  Black dots & High & $12$ & $24^{\circ}$\\  \hline
  Green diamonds & Low & $13.6$ & $29^{\circ}$\\  \hline
  Purple diamonds & Low & $23$ & $29^{\circ}$\\
   \hline
\end{tabular}
\caption{Data sets from Fig.~8 of Rubino \textit{et al.} \cite{Rubino2017}, which
are used in Fig.~5.}\label{tab:rubino}
\end{table}

In Sect.~IV, we also analyzed experimental data extracted from~\cite{Svetlizky2017a} in order to test the theoretical predictions in Eqs.~(6)-(7). The bulk parameters $\rho\=1170$ kg/m$^3$ and $c_s\=1345$ m/s, reported on in~\cite{Svetlizky2017a}, have been used together with $\mu\=\rho c_s^2$. The stress drops $\Delta\tau$ can be read off Fig.~2b of~\cite{Svetlizky2017a} (all figure indices in this and the next paragraphs refer to~\cite{Svetlizky2017a}), where the shear stress distribution near the edge of both slow (red) and rapid (here supershear, blue) rupture is presented. For rapid rupture, we extracted from Fig.~2b (blue data) $\Delta\tau\=1.17$ MPa. The corresponding particle velocity distribution $\dot{u}_x(x,y\=3.5\,\hbox{mm})$, measured $3.5$ mm {\em above} the interface, is presented in Fig.~1d (bottom). We used the leftmost value behind the edge, $\dot{u}_x(x,y\=3.5\,\hbox{mm})\!\simeq\!0.75$ m/s as an estimate for the tail particle velocity at the interface $\dot{u}_x(y\=0^+)$, from which we estimate the residual slip velocity to be $v_{\rm res}\=2\dot{u}_x(y\=0^+)\!\simeq\!1.5$ m/s. Note that the latter estimate is in very good agreement with the slip velocity reported for rapid (supershear) rupture in Fig.~3 (blue circles). Moreover, it is also in very good agreement with the normalized real area of contact $A_r/A_0\!\simeq\!0.6$, reported on in Fig.~1c, which according to Fig.~3b indeed corresponds to $v_{\rm res}$ slightly larger than $1$ m/s.

Using these estimates, we obtain
\begin{equation}
\frac{2c_s\Delta\tau}{\mu\,v_{\rm res}}=\frac{2\Delta\tau}{\rho\,c_s\,v_{\rm res}} = \frac{2 \times 1.17 \times 10^6}{1170 \times 1345 \times 1.5} = 0.99 \ ,
\end{equation}
in great agreement with the theoretical prediction Eq.~(5) in the manuscript. Repeating this procedure for slow rupture, we extracted from Fig.~2b (red data) $\Delta\tau\=0.242$ MPa and from Fig.~1d (top) $v_{\rm res}\=2\dot{u}_x(y\=0^+)\!\simeq\!2\times0.01=0.02$ m/s. Using these estimates, we obtain
\begin{equation}
\frac{2c_s\Delta\tau}{\mu\,v_{\rm res}}=\frac{2\Delta\tau}{\rho\,c_s\,v_{\rm res}} = \frac{2 \times 0.242 \times 10^6}{1170 \times 1345 \times 0.02} \simeq 15 \gg 1 \ ,
\end{equation}
in great agreement with the theoretical prediction Eq.~(7).

\section*{Appendix A7: Using benchmark crack solutions to estimate the integral contribution to the stress drop}
\label{sec:s_estimates}

We provide here details about the estimates of the spatiotemporal integral contribution to the stress drop presented in Sect.~IIB. As explained in Sect.~IIA, the relative magnitude of the latter, denoted by $s$, is given by $\delta v_{\rm res}/v^0_{\rm res}\=(v_{\rm res}-v^0_{\rm res})/v^0_{\rm res}$. Once $\Delta\tau$ is known, $v^0_{\rm res}$ is obtained through Eq.~(5). Our goal is to show how $v_{\rm res}$ can be estimated using benchmark crack solutions~\cite{Broberg1999Book}. Consider first anti-plane shear (mode-III symmetry) self-similar crack solutions, where the crack face displacement is given in Eq.~(6.9.145) in~\cite{Broberg1999Book} and takes the form
\begin{equation}
u_z(x,y\!=\!0^+,t)=\frac{\tau_{\rm d}\sqrt{(c_{\rm r}t)^2-x^2}}{\mu\,{\bm E}\big(\!1-c^2_{\rm r}/c^2_s\big)} \qquad\hbox{for}\qquad |x|<c_{\rm r} t \ ,
\label{eq:Brober_modeIII}
\end{equation}
where the notation has been adapted to be consistent with the present manuscript and ${\bm E}(\cdot)$ is the complete elliptic integral of the second kind~\cite{comment1}. Note that the center of the crack, whose length at time $t$ is $2c_{\rm r}t$, is located at $x\=0$.

In order to estimate $v_{\rm res}$ based on Eq.~\eqref{eq:Brober_modeIII}, follow these steps: (i) replace the far-field applied shear stress $\tau_{\rm d}$ by the stress drop $\Delta\tau$ to account for a finite frictional resistance, not included in the crack solution (ii) take the limit $c_{\rm r}\!\to\!c_s$ ($c_s$ is the limiting propagation speed in mode-III) because only in this limit the frictional rupture we consider mimics self-similar propagation (iii) focus on a region far behind the propagating tip, i.e.~set $x\!\to\!0$, which is relevant for the definition of a stress drop (iv) take the time derivative to obtain $\dot{u}_z(x\=0,y\=0^+,t)$ and finally (v) set $v_{\rm res}\=\dot{u}_z(x\=0,y\=0^+,t)-\dot{u}_z(x\=0,y\=0^-,t)\=2\dot{u}_z(x\=0,y\=0^+,t)$. The result reads
\begin{equation}
v_{\rm res}=\frac{2c_s\Delta\tau}{\mu\,{\bm E}(0)} \ .
\label{eq:Brober_modeIII_v_res}
\end{equation}
Using Eq.~(5), which implies $v^0_{\rm res}\=2c_s\Delta\tau/\mu$, together with Eq.~\eqref{eq:Brober_modeIII_v_res}, we obtain (as reported on in Sect.~IIB)
\begin{equation}
\delta v_{\rm res}/v^0_{\rm res}\=1/{\bm E}(0)-1=2/\pi-1\!\approx\!-0.36 \ ,
\label{eq:Brober_modeIII_delta_v_res}
\end{equation}
where ${\bm E}(0)\=\pi/2$ has been used.

In order to repeat this procedure for in-plane shear (mode-II symmetry) self-similar crack solutions, we start with the crack face displacement given in Eq.~(6.9.85) in~\cite{Broberg1999Book}
\begin{equation}
u_x(x,y\!=\!0^+,t)=\frac{\tau_{\rm d} (c_s^2-c_{\rm r}^2)\sqrt{(c_{\rm r} t)^2-x^2}}{\mu\,c_d^2\,g_2(c_{\rm r}/c_s,c_s/c_d)}
\label{eq:Brober_modeII}
\end{equation}
for $|x|\!<\!c_{\rm r} t$, where the function $g_2(\cdot)$ is given in Eq.~(6.9.87) in~\cite{Broberg1999Book} and $c_d$ is the dilatational wave speed. Considering the limit $c_{\rm r}\!\to\!c_R$ ($c_R$ is the Rayleigh wave-speed, the limiting speed for mode-II) and following the procedure described above, we obtain
\begin{equation}
v_{\rm res}=\frac{2c_R\Delta\tau}{\mu}\frac{\left(c_s^2-c_R^2\right)}{c_d^2\,g_2(c_R/c_s,c_s/c_d)} \ ,
\label{eq:Brober_modeII_delta_v_res}
\end{equation}
which leads to
\begin{equation}
\frac{\delta{v}_{\rm res}}{v_{\rm res}^0}=\frac{(c_s^2-c_R^2)\,c_R/c_s}{c_d^2\,g_2(c_R/c_s,c_s/c_d)}-1 \ .
\label{eq:Brober_modeII_delta_v_res}
\end{equation}
The latter is a function of $c_s/c_d$ (note that $c_R/c_s$ is also determined by $c_s/c_d$). Evaluating $g_2(c_R/c_s,c_s/c_d)$~\cite{comment2} for the whole range of admissible $c_s/c_d$ values (which in turn depends on Poisson's ratio) and plugging in Eq.~\eqref{eq:Brober_modeII_delta_v_res} shows that $\delta{v}_{\rm res}/v_{\rm res}^0$ varies between $-0.613$ and $-0.417$, as reported on in Sect.~IIB.

\end{document}